\documentclass[notitlepage,a4paper,aps,prd,onecolumn,superscriptaddress,nofootinbib,groupedaddress]{revtex4}

\usepackage{amsmath}
\usepackage{amsfonts,color}
\usepackage{amsmath}
\usepackage{amsthm}
\usepackage{pdfpages}
\usepackage{amsmath}
\usepackage{empheq}
\usepackage{amsfonts,color}
\usepackage{amssymb,float}
\usepackage{physics}
\usepackage{booktabs,multirow}
\usepackage{titlesec}

\titleformat{\paragraph}
{\normalfont\normalsize\bfseries}{\theparagraph}{1em}{}
\titlespacing*{\paragraph}
{0pt}{3.25ex plus 1ex minus .2ex}{1.5ex plus .2ex}

\usepackage{amsmath}
\usepackage[utf8]{inputenc}
\allowdisplaybreaks
\usepackage{color}
\usepackage{hyperref}
\hypersetup{
    colorlinks=true,
    linkcolor=blue,
    filecolor=magenta,      
   citecolor=blue
}

\usepackage{enumitem}

\usepackage{tikz}
\usetikzlibrary{shapes.geometric}
\usetikzlibrary{arrows.meta,arrows}

\begin{document}
\title{New black hole solutions with a dynamical traceless nonmetricity tensor in Metric-Affine Gravity}

\author{Sebastian Bahamonde}
\email{bahamonde.s.aa@m.titech.ac.jp}
\affiliation{Department of Physics, Tokyo Institute of Technology
1-12-1 Ookayama, Meguro-ku, Tokyo 152-8551, Japan.}
\affiliation{Laboratory of Theoretical Physics, Institute of Physics, University of Tartu, W. Ostwaldi 1, 50411 Tartu, Estonia.}

\author{Johann Chevrier}
\email{johann.chevrier@ens-paris-saclay.fr}
\affiliation{Département d'enseignement et de recherche de Physique, École Normale Supérieure de Paris-Saclay, Av. des Sciences 4, 91190 Gif-sur-Yvette, France.}

\author{Jorge Gigante Valcarcel}
\email{jorge.gigante.valcarcel@ut.ee}
\affiliation{Laboratory of Theoretical Physics, Institute of Physics, University of Tartu, W. Ostwaldi 1, 50411 Tartu, Estonia.}

\begin{abstract}

In the framework of Metric-Affine Gravity, the existing correspondence between the Einstein tensor and the energy-momentum tensor of matter provided by General Relativity is extended towards a post-Riemannian description in terms of the torsion and nonmetricity fields, which are sourced by the spin, dilation and shear currents of matter. In this work, we focus on the dynamical role of the traceless part of the nonmetricity tensor and its intrinsic connection with shears, defining a model which encloses a new black hole solution endowed with shear charges. We show that the extension in the presence of dynamical torsion and Weyl vector leads to the broadest family of static and spherically symmetric black hole solutions with spin, dilation and shear charges in Metric-Affine Gravity so far.

\end{abstract}

\maketitle

\section{Introduction}

Metric-Affine gauge theory of gravity, or Metric-Affine Gravity (MAG) for short, constitutes a natural extension of General Relativity (GR). Indeed, our current understanding of the gravitational interaction is based on the physical correspondence between the space-time curvature and the energy-momentum tensor of matter, as featured in the celebrated Einstein's field equations~\cite{Wald:1984rg}. In this regard, the framework of MAG incorporates the notions of torsion and nonmetricity in an enriched space-time geometry. In particular, the spin angular momentum of matter turns out to operate as a source of torsion~\cite{Hehl:1976kj,Obukhov:1987tz,Blagojevic:2013xpa,Obukhov:2022khx}, whereas the so-called dilation and shear currents of matter act as sources of nonmetricity~\cite{Hehl:1994ue,ponomarev2017gauge,Cabral:2020fax}.

Due to the lack of better insights and/or phenomenological evidence for the existence of these quantities, the formulation of new MAG models admitting exact solutions with torsion and nonmetricity beyond GR is especially relevant. In this sense, the main obstruction in finding such models is the computational intractability due to the simultaneous presence of curvature, torsion and nonmetricity in the calculations. Furthermore, the dynamical aspects of these post-Riemannian quantities depend on the order of their respective field strength tensors in the gravitational action: whereas the case with linear combinations of the curvature tensor involves nonpropagating torsion and nonmetricity tensors (i.e. tied to material sources with intrinsic hypermomentum)~\cite{Bakler:1985qj}, higher-order curvature corrections endow these fields with dynamics~\cite{Gronwald:1995em}, at the cost of considerably increasing the complexity of the theory. Nevertheless, substantial efforts to unravel the physical consequences of the theory, and more precisely, to find and scrutinize different types of solutions to the field equations, were made since its inception~\cite{adamowicz1980plane,Baekler:1981lkh,Gonner:1984rw,Bakler:1988nq,Tresguerres:1995js,tresguerres1995exact,Hehl:1999sb,Garcia:2000yi,Puetzfeld:2001hk}, whereas noteworthy applications in the fields of cosmology, black hole physics and gravitational waves have been more recently studied~\cite{Chen:2009at,Lu:2016bcx,Boos:2016cey,Cembranos:2016gdt,Cembranos:2017pcs,Blagojevic:2017wzf,Blagojevic:2017ssv,Obukhov:2019fti,Guerrero:2020azx,Bahamonde:2020fnq,Obukhov:2020hlp,Iosifidis:2020gth,Aoki:2020zqm,Bahamonde:2021qjk,delaCruzDombriz:2021nrg,Jimenez-Cano:2022arz,Bahamonde:2022meb,Bombacigno:2022lcx,Boudet:2022nub,Yang:2022icz,Iosifidis:2022xvp}.

Concerning black hole solutions with torsion and nonmetricity, the first results were initially found by considering a particularly simple form of the Lagrangian and further mathematical constraints on these tensors, which strongly
simplify the resulting field equations, such as the so-called triplet ansatz or other similar restrictions obtained via prolongation techniques~\cite{Hehl:1999sb,Heinicke:2005bp,Baekler:2006de}. In this regard, due to these strong simplifications the resulting black hole configurations do not show a full dynamical correspondence between the metric, the coframe and the corresponding
independent post-Riemannian structures present in the connection. Such a correspondence was recently obtained by the formulation of a new model endowed with dynamical torsion and nonmetricity in the framework of Weyl-Cartan geometry~\cite{Bahamonde:2020fnq}. Thereby, in the present work we address the extension of these results towards more general metric-affine geometries, where also the traceless part of the nonmetricity tensor becomes dynamical, which opens the door to further phenomenological analyses on this sector of the theory.

With this aim, we organise this paper as follows. In Sec.~\ref{sec:metricaffine}, we briefly introduce the foundations of MAG and the main geometrical features provided by a traceless nonmetricity tensor in the affine connection. In Sec.~\ref{sec:action}, we define a MAG model which includes higher-order curvature corrections displaying a dynamical traceless nonmetricity tensor, and we present the underlying field equations derived in the gauge formalism. The analysis and the resolution of these equations in a static and spherically symmetric space-time is carried out in Sec.~\ref{sec:BHtraceless}. Applying certain consistency constraints in Sec.~\ref{sec:BHtracelessA}, we find an exact Reissner-Nordstr\"{o}m-like solution with shear charges in Sec.~\ref{sec:BHtracelessB}. Furthermore, in Sec.~\ref{sec:fullmodel} we probe the compatibility of our results with the Reissner-Nordstr\"{o}m-like black holes previously found in~\cite{Bahamonde:2020fnq}, in the presence of dynamical torsion and Weyl vector, which leads to an extended gravitational action yielding solutions of this type with spin, dilation and shear charges. Finally, the main conclusions of our work are presented in Sec.~\ref{sec:conclusions}.

We work in natural units $c=G=1$, and we consider the metric signature $(+,-,-,-)$. In addition, we use a tilde accent to denote quantities associated with the general affine connection, whereas their unaccented counterparts are associated with the Levi-Civita connection. Latin and Greek indices run from $0$ to $3$, referring to anholonomic and coordinate bases, respectively.

\section{Metric-Affine Gravity in affinely connected metric space-times}\label{sec:metricaffine}

In the description of the gravitational interaction, the main feature of GR as the most successful and accurate theory of gravity is the existing correspondence between the geometry of the space-time and the energy-momentum properties of matter. From a more geometrical point of view, space-time is modelled as a smooth manifold equipped with a metric tensor $g_{\mu\nu}$ of Lorentzian signature, which induces a unique Levi-Civita affine connection $\Gamma^\lambda{}_{\mu\nu}$; providing the notions of distance and parallel transport, respectively. In fact, the metric tensor turns out to be the only dynamical variable of the theory even in the Palatini treatment of GR, where the metric tensor and the affine connection are initially considered as independent variables, in virtue of the projective invariance of its gravitational action~\cite{Dadhich:2010xa}. Thereby, the purely Riemannian character of the interaction does dynamically arise via a variational action principle, provided that matter does not couple to the general affine connection~\cite{BeltranJimenez:2017doy,Afonso:2017bxr,Sadovski:2022kwf}.

Nevertheless, considering the matter sector, the absence of finite spinor representations for the diffeomorphism group turns out to be supplied by a general principal bundle connection $\omega_{\mu} \in \mathfrak{so}(1,3)$, which allows the dynamics of spinor fields to be described into this geometrical scheme and leads to a gauge characterisation of the geometry of the space-time~\cite{Hehl:1976kj}. More generally, the consideration of an anholonomic connection $\omega_{\mu} \in \mathfrak{gl}(4,R)$ accounts for the spin, dilation and shear currents of matter, and fulfils via a vector bundle isomorphism, or coframe field $e^{a}\,_{\mu}$, the following fundamental relation with the affine connection of an affinely connected metric space-time~\cite{Hehl:1994ue}:
\begin{equation}\label{anholonomic_connection}
\omega^{a}\,_{b\mu}=e^{a}\,_{\lambda}\,e_{b}\,^{\rho}\,\tilde{\Gamma}^{\lambda}\,_{\rho \mu}+e^{a}\,_{\lambda}\,\partial_{\mu}\,e_{b}\,^{\lambda}\,,
\end{equation}
which is equipped with torsion and nonmetricity tensors
\begin{equation}
    T^{\lambda}\,_{\mu \nu}=2\tilde{\Gamma}^{\lambda}\,_{[\mu \nu]}\,, \quad Q_{\lambda \mu \nu}=\tilde{\nabla}_{\lambda}g_{\mu \nu}\,,
\end{equation}
where we can define the covariant derivative acting on an arbitrary tensor $T^{\lambda_{1}...\lambda_{i}}\,_{\rho_{1}...\rho_{j}}$ as
\begin{align}
\tilde{\nabla}_\mu T^{\lambda_{1}...\lambda_{i}}\,_{\rho_{1}...\rho_{j}}=&\;\partial_\mu T^{\lambda_{1}...\lambda_{i}}\,_{\rho_{1}...\rho_{j}}+\tilde{\Gamma}^{\lambda_{1}}{}_{\sigma\mu}T^{\sigma...\lambda_{i}}\,_{\rho_{1}...\rho_{j}}+...+\tilde{\Gamma}^{\lambda_{i}}{}_{\sigma\mu_{i}}T^{\lambda_{1}...\sigma}\,_{\rho_{1}...\rho_{j}}\nonumber\\
&-\tilde{\Gamma}^{\sigma}{}_{\rho_{1}\mu}T^{\lambda_{1}...\lambda_{i}}\,_{\sigma...\rho_{j}}-...-\tilde{\Gamma}^{\sigma}{}_{\rho_{j}\mu}T^{\lambda_{1}...\lambda_{i}}\,_{\rho_{1}...\sigma}\,. 
\end{align}

In that case, the affine connection displays a distortion part $N^{\lambda}\,_{\mu \nu}=\tilde{\Gamma}^{\lambda}\,_{\mu \nu}-\Gamma^{\lambda}\,_{\mu \nu}$, which incorporates the aforementioned tensors as follows:
\begin{equation}
    N^{\lambda}\,_{\mu\nu}=\frac{1}{2}\left(T^{\lambda}\,_{\mu \nu}-T_{\mu}\,^{\lambda}\,_{\nu}-T_{\nu}\,^{\lambda}\,_{\mu}\right)+\frac{1}{2}\left(Q^{\lambda}\,_{\mu \nu}-Q_{\mu}\,^{\lambda}\,_{\nu}-Q_{\nu}\,^{\lambda}\,_{\mu}\right)\,,
\end{equation}
and corrects the form of the curvature tensor
\begin{equation}\label{totalcurvature}
\tilde{R}^{\lambda}\,_{\rho\mu\nu}=R^{\lambda}\,_{\rho\mu\nu}+\nabla_{\mu}N^{\lambda}\,_{\rho \nu}-\nabla_{\nu}N^{\lambda}\,_{\rho \mu}+N^{\lambda}\,_{\sigma \mu}N^{\sigma}\,_{\rho \nu}-N^{\lambda}\,_{\sigma \nu}N^{\sigma}\,_{\rho \mu}\,.
\end{equation}

As is shown, the curvature tensor of an affinely connected metric space-time contains corrections provided by the presence of torsion and nonmetricity. Furthermore, the latter also leads to the definition of three independent traces of this tensor, namely the Ricci and co-Ricci tensors\footnote{Note that the trace of the Ricci and co-Ricci tensors provides a unique independent scalar curvature.}:
\begin{equation}
\tilde{R}_{\mu\nu}=\tilde{R}^{\lambda}\,_{\mu \lambda \nu}\,, \quad \hat{R}_{\mu\nu}=\tilde{R}_{\mu}\,^{\lambda}\,_{\nu\lambda}\,,
\end{equation}
as well as the so-called homothetic curvature tensor $\tilde{R}^{\lambda}\,_{\lambda\mu\nu}$, which encodes the change of lengths of vectors provided by the trace part of nonmetricity under their parallel transport along closed loops. Indeed, it is straightforward to check that the change of lengths of a given vector $k^{\mu}$ as well as the change of angles between two unit timelike vectors $\hat{m}^{\mu}$ and $\hat{n}^{\mu}$, under a general parallel transport defined by a tangent vector $V^{\mu}$, is proportional to the nonmetricity tensor:
\begin{align}
    V^{\lambda}\tilde{\nabla}_{\lambda}||\mathbf{k}||^{2}&=V^{\lambda}Q_{\lambda\mu\nu}k^{\mu}k^{\nu}\,,\label{change_lenghts}\\
    V^{\lambda}\tilde{\nabla}_{\lambda}\left(g_{\mu\nu}\hat{m}^{\mu}\hat{n}^{\nu}\right)&=V^{\lambda}Q_{\lambda\mu\nu}\hat{m}^{\mu}\hat{n}^{\nu}-\frac{1}{2}V^{\lambda}Q_{\lambda\mu\nu}\left(\hat{m}^{\mu}\hat{m}^{\nu}+\hat{n}^{\mu}\hat{n}^{\nu}\right)\hat{m}^{\rho}\hat{n}_{\rho}\label{change_angles}\,.
\end{align}
In particular, it is clear to note that Expression~\eqref{change_angles} vanishes identically when the nonmetricity tensor is totally ascribed to the so-called Weyl vector
\begin{equation}\label{Weyl_vector}
    W_{\mu}=\frac{1}{4}\,Q_{\mu\nu}\,^{\nu}\,.
\end{equation}

Therefore, the presence of a traceless part in the nonmetricity tensor alters the angles of vectors under parallel transport and does not preserve the light-cone structure. As a matter of fact, it is worthwhile to express nonmetricity in terms of the Weyl vector and a tensor ${\nearrow\!\!\!\!\!\!\!Q}_{\lambda\mu\nu}$, which is traceless in the indices $\mu$ and $\nu$, i.e. ${\nearrow\!\!\!\!\!\!\!Q}_{\lambda\mu}{}^{\mu}=0$\,:
\begin{equation}
Q_{\lambda\mu\nu}=g_{\mu\nu}W_{\lambda}+{\nearrow\!\!\!\!\!\!\!Q}_{\lambda\mu\nu}\,.
\end{equation}
The second piece is reducible, and further decomposes under the group of global Lorentz transformations~\cite{McCrea:1992wa} into:
\begin{equation}\label{irreducibletracelessnonmetricity}
    {\nearrow\!\!\!\!\!\!\!Q}_{\lambda\mu\nu}=g_{\lambda(\mu}\Lambda_{\nu)}-\frac{1}{4}g_{\mu\nu}\Lambda_{\lambda}+\frac{1}{3}\varepsilon_{\lambda\rho\sigma(\mu}\Omega_{\nu)}\,^{\rho\sigma}+q_{\lambda\mu\nu}\,,
\end{equation}
where $\varepsilon_{\lambda\rho\mu\nu}$ is the Levi-Civita (density) tensor and the pieces
\begin{align}
    \Lambda_{\mu}&=\frac{4}{9}\left(Q^{\nu}\,_{\mu\nu}-W_{\mu}\right)\,,\\
    \Omega_{\lambda}\,^{\mu\nu}&=-\,\left[\varepsilon^{\mu\nu\rho\sigma}Q_{\rho\sigma\lambda}+\varepsilon^{\mu\nu\rho}\,_{\lambda}\left(\frac{3}{4}\Lambda_{\rho}-W_{\rho}\right)\right]\,,\\
    q_{\lambda\mu\nu}&=Q_{(\lambda\mu\nu)}-g_{(\mu\nu}W_{\lambda)}-\frac{3}{4}g_{(\mu\nu}\Lambda_{\lambda)}\,,
\end{align}
constitute a vector, and two traceless and pseudotraceless tensors, respectively\footnote{We refer to ${\nearrow\!\!\!\!\!\!\!Q}_{\lambda\mu\nu}$ as the traceless nonmetricity tensor involving only the indices $\mu$ and $\nu$, whereas the irreducible tensor parts $\Omega_{\lambda}\,^{\mu\nu}$ and $q_{\lambda\mu\nu}$ are both traceless and pseudotraceless concerning all of their indices.}.

On the other hand, the main geometrical implication of the torsion tensor when considering the parallel transport of vectors is the breaking of standard parallelograms, being the resulting failure closed up only to a translation~\cite{Hehl:2007bn}. Likewise, it also presents its corresponding decomposition into irreducible modes~\cite{gambini1980einstein}:
\begin{equation}\label{irreducibletorsion}
T^{\lambda}\,_{\mu \nu}=\frac{1}{3}\left(\delta^{\lambda}\,_{\nu}T_{\mu}-\delta^{\lambda}\,_{\mu}T_{\nu}\right)+\frac{1}{6}\,\varepsilon^{\lambda}\,_{\rho\mu\nu}S^{\rho}+t^{\lambda}\,_{\mu \nu}\,,
\end{equation}
where
\begin{align}
T_{\mu}&=T^{\nu}\,_{\mu\nu}\,,\\
S_{\mu}&=\varepsilon_{\mu\lambda\rho\nu}T^{\lambda\rho\nu}\,,\\
t_{\lambda\mu\nu}&=T_{\lambda\mu\nu}-\frac{2}{3}g_{\lambda[\nu}T_{\mu]}-\frac{1}{6}\,\varepsilon_{\lambda\rho\mu\nu}S^{\rho}\,,
\end{align}
define a vector, an axial vector and a tensor with the same symmetry properties as $\Omega_{\lambda\mu\nu}$, respectively. 

From a phenomenological point of view, the different roles displayed by the aforementioned irreducible modes are significant, since they can be important in a large number of physically relevant processes and configurations. For instance, in the torsion sector, it is worthwhile to stress the interaction carried out by the axial vector with Dirac spinors under minimal coupling or the cosmological implications of both vector and axial parts within a spatially homogeneous and isotropic universe, where the additional tensor mode of torsion does not play any physical role under such a cosmological symmetry~\cite{Cembranos:2018ipn,tsamparlis1979cosmological}. For nonmetricity, the physical effects provided by a propagating Weyl vector in cosmology and black hole physics are well-known~\cite{Hehl:1999sb,Puetzfeld:2004yg,Bahamonde:2022meb}, whereas the dynamical implications of the additional traceless part remain largely unexplored and deserve further investigation. In this sense, one should expect a richer structure in the spectrum of solutions of MAG when considering its irreducible parts as dynamical fields, since they actually carry thirty six degrees of freedom of the forty nontrivial components of the nonmetricity tensor in four dimensions. Thereby, the consideration of a propagating traceless nonmetricity tensor in the gravitational action of MAG shall allow us to obtain a model where it is possible to find the broadest family of black hole solutions with dynamical torsion and nonmetricity in MAG.

\section{Gravitational action with a dynamical traceless nonmetricity tensor}\label{sec:action}

In order to introduce the dynamics of the torsion and nonmetricity fields, the most general parity conserving quadratic Lagrangian includes 11+3+4 irreducible modes from curvature, torsion and nonmetricity, respectively \cite{McCrea:1992wa}, which substantially increases the complexity of the theory. Therefore, the only way to tackle this mathematical issue and to shed light on the physical implications of torsion and nonmetricity has systematically required the study of particular models of MAG, where the coefficients of the Lagrangian are strongly restricted to a set of values, which turn out to describe different phenomenologies of these fields. Of course, such a procedure also aims to reach a broad model, constructed from each particular contribution and therefore allowing the description of the torsion and nonmetricity fields in a unified and clear way.

Following these lines, in the present case we first consider a quadratic parity preserving action displaying a dynamical traceless nonmetricity tensor as a geometrical correction to GR\footnote{Note that the Einstein-Hilbert term of the Lagrangian can be directly obtained from the general action of MAG, in virtue of the identity
\begin{eqnarray}
    \tilde{R}&=&R-2\nabla_{\mu}T^{\nu \mu}\,_{\nu}+\nabla_{\mu}Q^{\mu}\,_{\nu}\,^{\nu}-\nabla_{\mu}Q^{\nu}\,_{\nu}\,^{\mu}+\frac{1}{4}T_{\lambda \mu \nu}T^{\lambda \mu \nu}+\frac{1}{2}T_{\lambda \mu \nu}T^{\mu \lambda \nu}-T^{\lambda}\,_{\lambda\nu}T^{\mu}\,_{\mu}\,^{\nu}+T_{\lambda\mu\nu}Q^{\nu\lambda\mu}\nonumber\\
    &&+\,T^{\lambda}\,_{\lambda\nu}Q^{\mu\nu}\,_{\mu}-T^{\lambda}\,_{\lambda\nu}Q^{\nu\mu}\,_{\mu}+\frac{1}{4}Q_{\lambda\mu\nu}Q^{\lambda\mu\nu}-\frac{1}{2}Q_{\lambda\mu\nu}Q^{\mu\lambda\nu}+\frac{1}{2}Q^{\nu\lambda}\,_{\lambda}Q^{\mu}\,_{\mu\nu}-\frac{1}{4}Q^{\nu\lambda}\,_{\lambda}Q_{\nu}\,^{\mu}\,_{\mu}\,.
\end{eqnarray}}:
\begin{equation}\label{TracelessLagrangian}
S=\int
d^{4}x \sqrt{-g}\left\{\mathcal{L}_{\rm m}+\frac{1}{16\pi}\Bigl[
-\,R+2f_{1}\tilde{R}_{(\lambda\rho)\mu\nu}\tilde{R}^{(\lambda\rho)\mu\nu}+2f_{2}\left(\tilde{R}_{(\mu\nu)}-\hat{R}_{(\mu\nu)}\right)\left(\tilde{R}^{(\mu\nu)}-\hat{R}^{(\mu\nu)}\right)\Bigr]\right\}\,,
\end{equation}
where $\mathcal{L}_{\rm m}$ represents the matter Lagrangian. As can be seen, the propagation of the nonmetricity field described in the action is carried out by the symmetric part of the curvature tensor and its symmetric contraction
\begin{align}
    \tilde{R}^{\left(\lambda\rho\right)}\,_{\mu\nu}&=\tilde{\nabla}_{[\nu}Q_{\mu]}\,^{\lambda\rho}+\frac{1}{2}\,T^{\sigma}\,_{\mu\nu}Q_{\sigma}\,^{\lambda\rho}\,,\label{symmetric_curvature}\\
    \tilde{R}_{(\mu\nu)}-\hat{R}_{(\mu\nu)}&=\tilde\nabla_{(\mu}Q^\lambda{}_{\nu )\lambda}-\tilde\nabla_{\lambda}Q_{(\mu \nu )}{}^{\lambda}-Q^{\lambda\rho}\,_{\lambda}Q_{(\mu\nu)\rho}+Q_{\lambda\rho(\mu}Q_{\nu)}{}^{\lambda\rho}+T_{\lambda\rho(\mu}Q^{\lambda\rho}{}_{\nu)}\,,\label{symmetric_contraction}
\end{align}
which in turn constitute deviations from the third Bianchi of GR. By performing variations of Expression~\eqref{TracelessLagrangian} with respect to the coframe field and the anholonomic connection, the following field equations are derived:
\begin{eqnarray}
Y1_{\mu}\,^{\nu} &=& 8\pi\theta_{\mu}\,^{\nu}\label{traceless_field_eq1}\,,\\
\label{field_eq2}
Y2^{\lambda\mu\nu} &=& 4\pi\bigtriangleup^{\lambda\mu\nu}\label{traceless_field_eq2}\,,
\end{eqnarray}
where $Y1_{\mu}\,^{\nu}$ and $Y2^{\lambda\mu\nu}$ are tensor quantities defined in Appendix~\ref{sec:AppFieldEqs}, whereas $\theta_{\mu}\,^{\nu}$ and $\bigtriangleup^{\lambda\mu\nu}$ describe the canonical energy-momentum and hypermomentum density tensors of matter, respectively:
\begin{equation}
\theta_{\mu}\,^{\nu}=\frac{e^{a}\,_{\mu}}{\sqrt{- g}}\frac{\delta\left(\mathcal{L}_{m}\sqrt{- g}\right)}{\delta e^{a}\,_{\nu}}\,,
\end{equation}
\begin{equation}
\bigtriangleup^{\lambda\mu\nu}=\frac{e^{a\lambda}e_{b}\,^{\mu}}{\sqrt{- g}}\frac{\delta\left(\mathcal{L}_{m}\sqrt{-g}\right)}{\delta\omega^{a}\,_{b\nu}}\,.
\end{equation}

Therefore, both matter currents act as sources of the extended gravitational field. For general metric-affine geometries, the decomposition of the anholonomic connection (\ref{anholonomic_connection}) in the Lie algebra of the general linear group $GL(4,R)$ then leads hypermomentum to present its proper decomposition into spin, dilation and shear currents~\cite{Hehl:1994ue}. In particular, the presence of shears induces a dynamical traceless nonmetricity tensor, according to the internal structure of the special linear group $SL(4,R) \subset GL(4,R)$~\cite{Neeman:1978jlt}.

\section{Spherical symmetry and exact black hole solutions with shear charges}\label{sec:BHtraceless}

\subsection{Invariance conditions and consistency constraints}\label{sec:BHtracelessA}

The search of black hole configurations requires to solve the system of field equations~\eqref{traceless_field_eq1}-\eqref{traceless_field_eq2} in vacuum, which can be addressed by considering as a guiding principle the imposition of certain space-time symmetries. In the simplest case, the metric, torsion and nonmetricity tensors satisfy the same symmetry conditions provided by a Killing vector ${\xi}$ (see~\cite{Peterson:2019uzn} for alternative generalisations), which in turn are consequently reflected on the curvature tensor:
\begin{equation}
    \mathcal{L}_{\xi}g_{\mu\nu}=\mathcal{L}_{\xi}T^{\lambda}{}_{\mu\nu}=\mathcal{L}_{\xi}Q^{\lambda}{}_{\mu\nu}=0\,.
\end{equation}

In particular, for static and spherically symmetric space-times, these tensors are preserved under the action of the Killing vectors
\begin{align}
    \xi_{0}&=\partial_{t}\,,\\
    \xi_{1}&=\sin\varphi\,\partial_{\vartheta}+\cot\vartheta\cos\varphi\,\partial_{\varphi}\,,\\
    \xi_{2}&=-\,\cos\varphi\,\partial_{\vartheta}+\cot\vartheta\sin\varphi\,\partial_{\varphi}\,,\\
    \xi_{3}&=-\,\partial_{\varphi}\,,
\end{align}
which constrains their components as follows~\cite{Hohmann:2019fvf}:
\begin{align}\label{sph_metric}
    g_{tt}=\Psi_{1}(r)\,, \quad g_{rr}=-\,\frac{1}{\Psi_{2}(r)}\,, \quad g_{\vartheta\vartheta}=g_{\varphi\varphi}\csc^2\vartheta=-\,r^2\,,
\end{align}
\begin{align}\label{sph_torsion}
    T^t\,_{t r} &= t_{1}(r)\,, \quad T^r\,_{t r} = t_{2}(r)\,, \quad T^\vartheta\,_{t \vartheta} = T^\varphi\,_{t \varphi} = t_{3}(r)\,, \quad T^\vartheta\,_{r \vartheta} = T^\varphi\,_{r \varphi} =  t_{4}(r)\,, \\
    T^\vartheta\,_{t \varphi} &= T^\varphi\,_{\vartheta t} \sin^{2}\vartheta = t_{5}(r) \sin{\vartheta}\,, \quad T^\vartheta\,_{r \varphi} = T^\varphi\,_{\vartheta r} \sin^{2}\vartheta = t_{6}(r) \sin{\vartheta}\,, \\
    T^t\,_{\vartheta \varphi} &= t_{7}(r) \sin\vartheta\,, \quad T^r\,_{\vartheta \varphi} = t_{8}(r) \sin \vartheta\,,
\end{align}
\begin{align}\label{sph_nonmetricity}
    Q_{t t t} &= q_1(r)\,, \quad Q_{t r r}=q_2(r)\,, \quad Q_{t t r}= q_3(r)\,, \quad Q_{t \vartheta \vartheta}=Q_{t \varphi \varphi}\csc^2\vartheta=q_4(r)\,, \\
    Q_{r t t}&=q_5(r)\,, \quad Q_{r r r}=q_6(r)\,, \quad Q_{r t r}= q_7(r)\,, \quad Q_{r \vartheta \vartheta}=Q_{r \varphi \varphi}\csc^2\vartheta=q_8(r)\,, \\
    Q_{\vartheta t \vartheta}&= Q_{\varphi t \varphi}\csc^2\vartheta =  q_9(r)\,, \quad
    Q_{\vartheta r \vartheta}= Q_{\varphi r \varphi}\csc^2\vartheta = q_{10}(r)\,, \\
    Q_{\vartheta t \varphi}&= -\,Q_{\varphi t \vartheta} = q_{11}(r) \sin \vartheta\,, \quad Q_{\vartheta r \varphi}= -\,Q_{\varphi r \vartheta} = q_{12}(r) \sin \vartheta\,,
\end{align}
where $(t,r,\vartheta,\varphi)$ denote spherical coordinates. The tetrad field can then be written in the orthonormal gauge as
\begin{equation}\label{eq:Diagonal tetrad}
    e^a\,_\mu=\textrm{diag}\left(\sqrt{\Psi_{1}(r)},\frac{1}{\sqrt{\Psi_{2}(r)}},r,r\sin\vartheta\right).
\end{equation}

Thereby, the static and spherically symmetric  metric, torsion and nonmetricity tensors are described by two, eight and twelve independent functions of the radial coordinate, respectively. Hence, the large number of degrees of freedom present in the geometry and the highly nonlinear character of the field equations require the assumption of additional constraints, in order to find exact black hole solutions.

First, it is straightforward to note the presence of two degrees of freedom which correspond to the Weyl vector. In fact, it is clear from Expression~\eqref{symmetric_curvature} that the present model displays a kinetic term $f_{1}F^{(W)}_{\mu\nu}F^{(W)\mu\nu}/(8\pi)$ in the Lagrangian, with
\begin{equation}
    F^{(W)}_{\mu\nu}=2\partial_{[\mu}W_{\nu]}\,,
\end{equation}
which means that vanishing the Weyl vector switches off the homothetic part of the curvature tensor, associated with dilations. Thus, for simplicity we can initially neglect these components from our analysis, since they do not describe any traceless part of the nonmetricity tensor, giving rise to the expressions
\begin{eqnarray}\label{cond1}
    q_{1}(r)= \frac{\Psi_{1}(r)}{r^2}\left(r^{2}q_{2}(r)\Psi_{2}(r)+2q_{4}(r)\right)\,,\quad q_{5}(r)=\frac{\Psi_{1}(r)}{r^2} \left(r^2 q_6(r) \Psi_{2}(r)+2 q_{8}(r)\right)\,.
\end{eqnarray}

On the other hand, the realisation of shear transformations in the general linear group involves the part of the anholonomic connection describing a shear current or charge to take values in the symmetric traceless part of the Lie algebra $\mathfrak{gl}(4,R)$, which in virtue of Expression~\eqref{anholonomic_connection} is equivalent to the constraint
\begin{equation}\label{antisim_distortion}
    N_{[\lambda\rho]\mu}=0\,,
\end{equation}
where
\begin{equation}
    N_{[\lambda\rho]\mu}=\frac{2}{3}g_{\mu[\lambda}T_{\rho]}+\frac{3}{4}g_{\mu[\lambda}\Lambda_{\rho]}+\frac{1}{12}\varepsilon_{\lambda\rho\mu\sigma}S^{\sigma}+\frac{1}{2}\left(2t_{[\lambda\rho]\mu}-t_{\mu\lambda\rho}\right)-\frac{1}{6}\left(\Omega_{[\lambda}\,^{\sigma\nu}\varepsilon_{\rho]\sigma\nu\mu} -\Omega_{\mu}\,^{\sigma\nu}\varepsilon_{\lambda\rho\sigma\nu}\right)\,.
\end{equation}
In terms of the torsion and nonmetricity tensors, this constraint means $T_{\lambda\mu\nu}=Q_{[\mu\nu]\lambda}$, which sets distortion as
\begin{equation}
    N_{\lambda\rho\mu}=-\,\frac{1}{2}Q_{\mu\lambda\rho}\,.
\end{equation}
Indeed, it gives rise to the following relations between the irreducible modes of torsion and nonmetricity:
\begin{equation}
   T_{\mu}=-\,\frac{9}{8}\Lambda_{\mu}\,, \quad S_{\mu}=0\,, \quad t_{\lambda\mu\nu}=\frac{1}{4}\varepsilon_{\sigma\rho\mu\nu}\Omega_\lambda{}^{\sigma\rho}\,,
\end{equation}
and furthermore it reduces the expressions of the field strength tensors of the model to covariant derivatives of the Levi-Civita connection acting on the irreducible modes of the traceless nonmetricity tensor
\begin{align}
    \tilde{R}^{(\lambda\rho)}\,_{\mu\nu}&=\frac{1}{8}\,g^{\lambda \rho}F^{(\Lambda)}_{\mu\nu}-\nabla_{[\mu}\Lambda^{(\lambda}\delta_{\nu]}^{\rho)} + \frac{1}{3}\,\varepsilon_{\alpha\beta[\mu}\,^{(\lambda}\nabla_{\nu]}\Omega^{\rho)\alpha\beta}-\nabla_{[\mu}q^{\lambda \rho}\,_{\nu]}\,,\label{fs1}\\
    \tilde{R}_{(\mu\nu)}-\hat{R}_{(\mu\nu)}&=2\nabla_{(\mu}\Lambda_{\nu)}-\frac{1}{2}\,g_{\mu\nu}\nabla_{\lambda}\Lambda^{\lambda}+\frac{1}{6}\,\varepsilon^{\sigma}\,_{\lambda\rho(\mu|}\nabla_{\sigma}\Omega_{|\nu)}\,^{\lambda\rho}-\nabla_{\lambda}q^{\lambda}\,_{\mu\nu}\,,\label{fs2}
\end{align}
with
\begin{equation}
    F^{(\Lambda)}_{\mu\nu}=2\partial_{[\mu}\Lambda_{\nu]}\,.
\end{equation}
It is worthwhile to note these implications hold within any affinely connected metric space-time (i.e. not only in the static and spherically symmetric case). In any case, for a static and spherically symmetric space-time, the constraint~\eqref{antisim_distortion} is satisfied if
\begin{eqnarray}\label{cond2a}
    q_2(r)&=&q_7(r)-\frac{2 t_{2}(r)}{\Psi_{2}(r)}\,,\quad q_3(r)=\Psi_{1}(r) \left(2t_{1}(r)+q_6(r) \Psi_{2}(r)+\frac{2q_8(r)}{r^2}\right)\,,\\
    q_4(r)&=&q_9(r)-2r^2t_{3}(r)\,,\quad q_{10}(r)=q_8(r)+2r^2t_{4}(r)\,,\quad q_{11}(r)=t_{7}(r)\Psi_{1}(r)\,,\\ q_{12}(r)&=&-\,\frac{t_{8}(r)}{\Psi_{2}(r)}\,,\quad  t_{5}(r)=-\,\frac{t_{7}(r)\Psi_{1}(r)}{2r^2}\,,\quad t_{6}(r)=\frac{t_{8}(r)}{2 r^2\Psi_{2}(r)}\,.\label{cond2b}
\end{eqnarray}

Finally, we demand the corresponding torsion and nonmetricity scalars of the solution to be regular. This condition can be directly fixed if the expressions of the torsion and nonmetricity fields referred to the boosted basis\footnote{For simplicity, we omit the radial dependence in the functions.}
\begin{eqnarray}
\vartheta^{\hat{0}}&=&\frac{1}{2}{\Big(\frac{\Psi_1}{ \Psi_2}\Big)^{1/4}}\left\{ \left[\sqrt{\Psi_{1}\Psi_{2}}+1\right]\,dt+\left[1-\frac{1}{\sqrt{\Psi_{1}\Psi_{2}}}\right]\,dr \right\}\,,\\
\vartheta^{\hat{1}}&=&\frac{1}{2}{\Big(\frac{\Psi_1}{ \Psi_2}\Big)^{1/4}}\left\{ \left[\sqrt{\Psi_{1}\Psi_{2}}-1\right]\,dt+\left[1+\frac{1}{\sqrt{\Psi_{1}\Psi_{2}}}\right]\,dr \right\}\,,\\
\vartheta^{\hat{2}}&=&r\,d \vartheta \,,\\%
\vartheta^{\hat{3}}&=&r\sin\vartheta \, d \varphi\,,
\end{eqnarray}
do not present any explicit singular term in the roots of the metric functions $\Psi_{1}(r)$ and $\Psi_{2}(r)$. 
Specifically, the expression of the torsion field $\mathcal{F}^{a}\,_{b c} = \vartheta^{a}\,_{\lambda}\vartheta_{b}\,^{\mu}\vartheta_{c}\,^{\nu}T^{\lambda}\,_{\nu\mu}\,$ evaluated in the aforementioned rotated frame reads
\begin{eqnarray}
\mathcal{F}^{\hat{0}}\,_{\hat{0}\hat{1}} &=&-\color{black}\,{\frac{1}{2}\Big(\frac{\Psi_2}{\Psi_1}\Big)^{1/4}}\left\{\left[1+\sqrt{\Psi_{1}\Psi_{2}}\right]t_{1}+\left[1-\frac{1}{\sqrt{\Psi_{1}\Psi_{2}}}\right]t_{2}\right\}\,,\\
\mathcal{F}^{\hat{1}}\,_{\hat{0}\hat{1}} &=&-\color{black}\,{\frac{1}{2}\Big(\frac{\Psi_2}{\Psi_1}\Big)^{1/4}}\left\{\left[1+\frac{1}{\sqrt{\Psi_{1}\Psi_{2}}}\right]t_{2}-\left[1-\sqrt{\Psi_{1}\Psi_{2}}\right]t_{1}\right\} \,,\\
\mathcal{F}^{\hat{2}}\,_{\hat{0}\hat{2}} &=&\mathcal{F}^{\hat{3}}\,_{\hat{0}\hat{3}} =-\color{black}\,{\frac{1}{2}\Big(\frac{\Psi_2}{\Psi_1}\Big)^{1/4}}\left\{\left[1+\frac{1}{\sqrt{\Psi_{1}\Psi_{2}}}\right]t_{3}+\left[1-\sqrt{\Psi_{1}\Psi_{2}}\right]t_{4}\right\}\,,\\
\mathcal{F}^{\hat{2}}\,_{\hat{1}\hat{2}} &=& \mathcal{F}^{\hat{3}}\,_{\hat{1}\hat{3}} =-\color{black}\,{\frac{1}{2}\Big(\frac{\Psi_2}{\Psi_1}\Big)^{1/4}}\left\{\left[1+\sqrt{\Psi_{1}\Psi_{2}}\right]t_{4}-\left[1-\frac{1}{\sqrt{\Psi_{1}\Psi_{2}}}\right]t_{3}\right\} \,,\\
\mathcal{F}^{\hat{0}}\,_{\hat{2} \hat{3}} &=& 2\mathcal{F}^{\hat{3}}\,_{\hat{0} \hat{2}} = -2\mathcal{F}^{\hat{2}}\,_{\hat{0} \hat{3}}=-\color{black} \,{\frac{1}{2r^2}\Big(\frac{\Psi_1}{\Psi_2}\Big)^{1/4}}\left\{\left[1+\sqrt{\Psi_{1}\Psi_{2}}\right]t_{7}+\left[1-\frac{1}{\sqrt{\Psi_{1}\Psi_{2}}}\right]t_{8}\right\} \,,\label{torsionformaa}\\
\mathcal{F}^{\hat{1}}\,_{\hat{2} \hat{3}} &=&2\mathcal{F}^{\hat{2}}\,_{\hat{1} \hat{3}} = -\,2\mathcal{F}^{\hat{3}}\,_{\hat{1} \hat{2}}= -\,{\frac{1}{2r^2}\Big(\frac{\Psi_1}{\Psi_2}\Big)^{1/4}}\left\{\left[1+\frac{1}{\sqrt{\Psi_{1}\Psi_{2}}}\right]t_{8}-\left[1-\sqrt{\Psi_{1}\Psi_{2}}\right]t_{7}\right\} \,.\label{torsionformab}
\end{eqnarray}
Thereby, it is straightforward to note that the explicit divergent terms of the torsion field are cancelled out if
\begin{eqnarray}\label{rel1_tors}
     t_{2}(r)=t_{1}(r)\sqrt{\Psi_{1}(r)\Psi_{2}(r)}\,,\quad  t_{3}(r)=-\,t_{4}(r)\sqrt{\Psi_{1}(r)\Psi_{2}(r)}\,,\quad  t_{8}(r)=t_{7}(r)\sqrt{\Psi_{1}(r)\Psi_{2}(r)}\,,
\end{eqnarray}
which actually implies that the corresponding torsion scalars are regular and equal to zero.

Applying the same reasoning for the rotated nonmetricity field $\mathcal{G}_{a b c} = \vartheta_{a}\,^{\mu}\vartheta_{b}\,^{\nu}\vartheta_{c}\,^{\lambda}Q_{\lambda\mu\nu}\,$, one finds the components
\begin{eqnarray}
    \mathcal{G}_{\hat{0}\hat{0}\hat{0}}&=&\frac{1}{4r^2}\Big(\frac{\Psi_1}{\Psi_2}\Big)^{1/4}\biggl\{\frac{2 r^2 (q_6 \Psi_2+t_4)+3 q_8+3 q_9}{\Psi_1}-\Psi_2^3 \sqrt{\Psi_1 \Psi_2} \left[2 r^2 (q_6 \Psi_2+2 t_1-t_4)+3 q_8\right]\nonumber\\
    &&+\Psi_2 \left[2 r^2 (q_7 \Psi_2-2 t_1+3 t_4)-3 q_8+q_9\right]+\frac{2 r^2 q_7 \Psi_2+q_9}{\Psi_1^{3} \sqrt{\Psi_2\Psi_1}}+3 \sqrt{\frac{\Psi_2}{\Psi_1}} \left(q_8+q_9+2 r^2 t_4\right)\biggr\}\,,\\
    \mathcal{G}_{\hat{0}\hat{1}\hat{0}}&=&\frac{1}{4r^2}\Big(\frac{\Psi_1}{\Psi_2}\Big)^{1/4}\biggl\{\frac{2 r^2 (q_6 \Psi_2+t_4)+3 q_8+q_9}{\Psi_1}+\Psi_2^3 \sqrt{\Psi_1 \Psi_2} \left(2 r^2 (q_6 \Psi_2+2 t_1-t_4)+3 q_8\right)\nonumber\\
    &&+\Psi_2 \left(2 r^2 (-q_7 \Psi_2+2 t_1-t_4)+q_8-q_9\right)+\frac{2 r^2 q_7 \Psi_2+q_9}{\Psi_1^{3/2} \sqrt{\Psi_2}}+\sqrt{\frac{\Psi_2}{\Psi_1}} \left(q_8-q_9+2 r^2 t_4\right)\biggr\}\,,\\
    \mathcal{G}_{\hat{1}\hat{1}\hat{0}}&=&\frac{1}{4r^2}\Big(\frac{\Psi_1}{\Psi_2}\Big)^{1/4}\biggl\{\frac{2 r^2 (q_6 \Psi_2+t_4)+3 q_8-q_9}{\Psi_1}-\Psi_2^3 \sqrt{\Psi_1 \Psi_2} \left(2 r^2 (q_6 \Psi_2+2 t_1-t_4)+3 q_8\right)\nonumber\\
    &&+\Psi_2 \left(-2 r^2 (-q_7 \Psi_2+2 t_1+t_4)+q_8+q_9\right)+\frac{2 r^2 q_7 \Psi_2+q_9}{\Psi_1^{3/2} \sqrt{\Psi_2}}-\sqrt{\frac{\Psi_2}{\Psi_1}} \left(q_8+q_9+2 r^2 t_4\right)\biggr\}\,,\\
    \mathcal{G}_{\hat{0}\hat{0}\hat{1}}&=&\frac{1}{4r^2}\Big(\frac{\Psi_1}{\Psi_2}\Big)^{1/4}\biggl\{\frac{2 r^2 (q_6 \Psi_2+t_4)+3 q_8+q_9}{\Psi_1}+\Psi_2^3 \sqrt{\Psi_1 \Psi_2} \left(2 r^2 (q_6 \Psi_2+2 t_1-t_4)+3 q_8\right)\nonumber\\
    &&-\Psi_2 \left(2 r^2 (q_7 \Psi_2+2 t_1+t_4)-q_8+q_9\right)+\frac{2 r^2 q_7 \Psi_2+q_9}{\Psi_1^{3/2} \sqrt{\Psi_2}}+\sqrt{\frac{\Psi_2}{\Psi_1}} \left(q_8-q_9+2 r^2 t_4\right)\biggr\}\,,\\
    \mathcal{G}_{\hat{0}\hat{1}\hat{1}}&=&\frac{1}{4r^2}\Big(\frac{\Psi_1}{\Psi_2}\Big)^{1/4}\biggl\{\frac{2 r^2 (q_6 \Psi_2+t_4)+3 q_8-q_9}{\Psi_1}-\Psi_2^3 \sqrt{\Psi_1 \Psi_2} \left(2 r^2 (q_6 \Psi_2+2 t_1-t_4)+3 q_8\right)\nonumber\\
    &&+\Psi_2 \left(2 r^2 (q_7 \Psi_2+2 t_1-t_4)+q_8+q_9\right)+\frac{2 r^2 q_7 \Psi_2+q_9}{\Psi_1^{3/2} \sqrt{\Psi_2}}-\sqrt{\frac{\Psi_2}{\Psi_1}} \left(q_8+q_9+2 r^2 t_4\right)\biggr\}\,,\\
    \mathcal{G}_{\hat{1}\hat{1}\hat{1}}&=&\frac{1}{4r^2}\Big(\frac{\Psi_1}{\Psi_2}\Big)^{1/4}\biggl\{\frac{2 r^2 (q_6 \Psi_2+t_4)+3 q_8-3 q_9}{\Psi_1}+\Psi_2^3 \sqrt{\Psi_1 \Psi_2} \left(2 r^2 (q_6 \Psi_2+2 t_1-t_4)+3 q_8\right)\nonumber\\
    &&-\Psi_2 \left(2 r^2 (q_7 \Psi_2+2 t_1-3 t_4)+3 q_8+q_9\right)+\frac{2 r^2 q_7 \Psi_2+q_9}{\Psi_1^{3/2} \sqrt{\Psi_2}}-3 \sqrt{\frac{\Psi_2}{\Psi_1}} \left(q_8-q_9+2 r^2 t_4\right)\biggr\}\,,\\
    \mathcal{G}_{\hat{2}\hat{2}\hat{0}}&=&\mathcal{G}_{\hat{3}\hat{3}\hat{0}}=\frac{1}{2r^2}\Big(\frac{\Psi_1}{\Psi_2}\Big)^{1/4}\biggl\{\sqrt{\frac{\Psi_2}{\Psi_1}}\left[q_8+2r^2t_4+\frac{q_9}{\sqrt{\Psi_1\Psi_2}}\right]-\Psi_2 \left[q_8-2r^2t_4-\frac{q_9}{\sqrt{\Psi_1\Psi_2}}\right]\biggr\}\,,\\
    \mathcal{G}_{\hat{2}\hat{2}\hat{1}}&=&\mathcal{G}_{\hat{3}\hat{3}\hat{1}}=\frac{1}{2r^2}\Big(\frac{\Psi_1}{\Psi_2}\Big)^{1/4}\biggl\{\sqrt{\frac{\Psi_2}{\Psi_1}}\left[q_8+2 r^2t_4+\frac{q_9}{\sqrt{\Psi_1\Psi_2}}\right]+\Psi_2\left[q_8-2 r^2 t_4-\frac{q_9}{\sqrt{\Psi_1\Psi_2}}\right]\biggr\}\,,\\
    \mathcal{G}_{\hat{0}\hat{2}\hat{2}}&=&\mathcal{G}_{\hat{0}\hat{3}\hat{3}}=\frac{1}{2r^2}\Big(\frac{\Psi_1}{\Psi_2}\Big)^{1/4}\biggl\{\sqrt{\frac{\Psi_2}{\Psi_1}}\left[q_8+2 r^2 t_4+\frac{q_9}{\sqrt{\Psi_1\Psi_2}}\right]-\Psi_2 \left[q_8+2 r^2 t_4-\frac{q_9}{\sqrt{\Psi_1\Psi_2}}\right]\biggr\}\,,\\
    \mathcal{G}_{\hat{1}\hat{2}\hat{2}}&=&\mathcal{G}_{\hat{1}\hat{3}\hat{3}}=\frac{1}{2r^2}\Big(\frac{\Psi_1}{\Psi_2}\Big)^{1/4}\biggl\{\sqrt{\frac{\Psi_2}{\Psi_1}}\left[q_8+2 r^2 t_4+\frac{q_9}{\sqrt{\Psi_1\Psi_2}}\right]+\Psi_2 \left[q_8+2 r^2 t_4-\frac{q_9}{\sqrt{\Psi_1\Psi_2}}\right]\biggr\}\,,\\
    \mathcal{G}_{\hat{0}\hat{2}\hat{3}}&=&-\,\mathcal{G}_{\hat{0}\hat{3}\hat{2}}=-\,\mathcal{G}_{\hat{1}\hat{2}\hat{3}}=\mathcal{G}_{\hat{1}\hat{3}\hat{2}}= -\,\frac{t_{7}\Psi_{1}^{3/4}\Psi_{2}^{1/4}}{r^2}\,,
\end{eqnarray}
in such a way that the explicit singularities in the roots of the functions $\Psi_1(r)$ and $\Psi_2(r)$ are removed if
\begin{eqnarray}
     q_6(r)&=&-\,\frac{q_7(r)}{\sqrt{\Psi_1(r)\Psi_2(r)}}-\frac{q_8(r)}{r^{2}\Psi_2(r)}\,,\label{cond3a}\\
     q_9(r)&=&-\,\sqrt{\Psi_{1}(r)\Psi_{2}(r)}\left(q_8(r)+2r^2t_{4}(r)\right)\,,\label{cond3b}
\end{eqnarray}
which in turn vanishes the respective nonmetricity scalars.

The application of these consistency constraints sets the metric, torsion and nonmetricity functions $\Psi_{1}(r)$, $\Psi_2(r)$, $t_1(r)$, $t_4(r)$, $t_7(r)$, $q_7(r)$ and $q_8(r)$ as the remaining unknown quantities of the model, and strongly simplifies the form of the field equations. This result turns out to be crucial for their resolution, since it allows us to solve all the field equations of the model in a systematic way, as shown in the following section.

\subsection{Resolution of the field equations}\label{sec:BHtracelessB}

This section will be devoted to solving the field equations of the theory, namely the tetrad and connection field equations~\eqref{tetradeqcase1} and~\eqref{connectioneqcase1}. For this task, we will focus on the connection equations to find the torsion and nonmetricity functions which solve them, whereas the tetrad equations will be used to determine the metric functions.

Concerning the different values of the Lagrangian coefficients, it is worthwhile to stress the existence of two main branches: (a) $f_1=0$  and (b) $f_1\neq0$. For the second main branch, it is easy to show that the field equations imply $t_7(r)=0$. This can be seen by solving the connection field equations $Y2^{\vartheta \varphi t}=0$ and $Y2^{\vartheta t \varphi}=0$ for $t_1(r)$ and $t_4(r)$, respectively. After doing this and replacing their expressions into the equation $Y2^{\varphi r \vartheta}=0$, one finds that the condition $t_7(r)=0$ must hold. Furthermore, by isolating $t_1'(r)$ from $Y2^{\varphi\varphi r}=0$ and replacing its expression into the component $Y2^{\varphi\varphi t}$, one finds two different cases within (b): i) $f_2=-f_1/2$ and ii) $f_2\neq -f_1/2$. We will demonstrate that a dynamical Reissner-Nordstr\"{o}m-like solution arises in (b)-(ii) if $f_{2}=-\,f_{1}/4$, the latter being the only combination of the Lagrangian coefficients for which the derivatives of $\Lambda_\mu$ appear in a gauge-consistent fashion, viz, via the field strength tensor $F^{(\Lambda)}_{\mu\nu}=2\partial_{[\mu}\Lambda_{\nu]}$. This can be shown by performing a post-Riemannian expansion of the Lagrangian presented in Expression~\eqref{TracelessLagrangian}, and isolating the part containing only the modes included in ${\nearrow\!\!\!\!\!\!\!Q_{\lambda\mu\nu}}$, namely
\begin{eqnarray}
    16\pi{\nearrow\!\!\!\!\!\!\!\mathcal{L}}_{\mathrm{kin}}&=&2\,(f_1+4f_2)\nabla_\mu \Lambda_\nu \nabla^\mu\Lambda^\nu-\frac{1}{2}\left( {f_1}+4f_2\right)\nabla_{\mu}\Lambda^{\mu}\nabla_{\nu}\Lambda^{\nu}-\frac{1}{8}\left(f_{1}+16f_{2}\right)F^{(\Lambda)}_{\mu\nu}F^{(\Lambda)\mu\nu}\nonumber\\
    &&-\,\frac{f_1}{3}{F}_{\mu\nu}^{(\Lambda)}\nabla_\lambda {\ast\Omega}^{\lambda\mu\nu}-2(f_1+4f_2)\nabla_{\mu}\Lambda_{\nu}\nabla_\lambda q^{\lambda\mu\nu}+\frac{4f_2}{3} \nabla^\mu {*\Omega}_{\lambda\rho\mu}\nabla_\nu q^{\nu\lambda\rho}-\frac{2f_1}{3}\nabla^\mu\Lambda^\nu\nabla^\lambda {*\Omega}_{\nu\mu\lambda}\nonumber\\
    &&-\,\frac{8f_2}{3}\nabla^\mu \Lambda^\nu \nabla^\lambda  {*\Omega}_{(\mu\nu)\lambda}+\frac{4f_1}{3}\nabla_\mu {*\Omega}_{\lambda\nu\rho}\nabla^\rho q^{\mu\lambda\nu}-\frac{4f_1}{9}\nabla_\lambda\Omega^{\mu\nu\lambda}\nabla^\rho\Omega_{\mu\nu\rho}+\frac{f_2}{18}\nabla_\lambda\Omega^{\lambda\mu\nu}\nabla^\rho \Omega_{\rho\mu\nu}\nonumber\\
    &&-\,\frac{f_1}{9}  \nabla_{\rho }\Omega_{\mu \lambda\nu } \nabla^{\mu }\Omega^{\rho  \lambda \nu} + \frac{2f_2}{9}  \nabla_{\lambda }\Omega_{\mu\nu \rho } \nabla^{\rho }\Omega^{\mu\nu \lambda } -\frac{1}{9} ( f_{1}{} +  f_{2}{}) \nabla_{\rho }\Omega_{\lambda\mu\nu} \nabla^{\rho }\Omega^{\lambda\mu\nu}\nonumber\\
    &&+\,2 f_{2}{} \nabla_{\lambda }q^{\lambda\mu\nu } \nabla^{\rho }q_{\rho\mu\nu} -  f_{1}{} \nabla_{\lambda }q_{\mu\nu \rho } \nabla^{\rho }q^{\mu\nu \lambda } + f_{1}{} \nabla_{\rho }q_{\lambda\mu\nu } \nabla^{\rho }q^{\lambda\mu\nu}\,,
\end{eqnarray}
with $*\Omega_{\lambda\mu\nu}=(1/2)\,\varepsilon^{\rho\sigma}{}_{\mu\nu}\Omega_{\lambda\rho\sigma}$. Indeed, for $f_2=-f_1/4$, we have 
\begin{eqnarray}\label{model_expansion}
    16\pi{\nearrow\!\!\!\!\!\!\!\mathcal{L}}_{\mathrm{kin}}&=&f_{1}\left(\frac{3}{8}F^{(\Lambda)}_{\mu\nu}F^{(\Lambda)\mu\nu}-\frac{1}{2}{F}^{(\Lambda)}_{\mu\nu}\nabla^\lambda {*\Omega}_{\lambda}\,^{\mu\nu}-\frac{1}{3} \nabla^\mu{*\Omega}_{\lambda\rho\mu}\nabla_\nu q^{\nu\lambda\rho}+\frac{4}{3}\nabla_\mu {*\Omega}_{\lambda\nu\rho}\nabla^\rho q^{\mu\lambda\nu}
    \right.
    \nonumber\\
    & &
    \left.
    -\,\frac{4}{9}\nabla_\lambda\Omega^{\mu\nu\lambda}\nabla^\rho\Omega_{\mu\nu\rho}-\frac{1}{72}\nabla_\lambda\Omega^{\lambda\mu\nu}\nabla^\rho \Omega_{\rho\mu\nu}-\,\frac{1}{9}  \nabla_{\rho }\Omega_{\mu \lambda\nu } \nabla^{\mu }\Omega^{\rho  \lambda \nu} - \frac{1}{18}  \nabla_{\lambda }\Omega_{\mu\nu \rho } \nabla^{\rho }\Omega^{\mu\nu \lambda }
    \right.
    \nonumber\\
    & &
    \left.
    -\,\frac{1}{12} \nabla_{\rho }\Omega_{\lambda\mu\nu } \nabla^{\rho }\Omega^{\lambda\mu\nu }-\frac{1}{2} \nabla_{\lambda }q^{\lambda\mu\nu } \nabla^{\rho }q_{\rho\mu\nu} -   \nabla_{\lambda }q_{\mu\nu \rho } \nabla^{\rho }q^{\mu\nu \lambda } +  \nabla_{\rho }q_{\lambda\mu\nu } \nabla^{\rho }q^{\lambda\mu\nu }
    \right),
\end{eqnarray}
where we have used the dimension-dependent identity $\ast\Omega_{[\lambda\mu\nu]}=0$, from which it follows that
\begin{equation}
    {\ast \Omega}_{[\mu\nu]\lambda}=-\,\frac{1}{2}{\ast\Omega}_{\lambda\mu\nu}\,.
\end{equation}

Following these lines, the analysis of the field equations for the two main branches of the model is shown below.

\subsubsection{First main branch ($f_1=0$)}

Let us now analyse the first main branch where $f_1=0$, which switches off the field strength tensor~\eqref{symmetric_curvature} from the Lagrangian. In that case, we can easily solve~$Y2^{\vartheta \varphi t}=0$ for $t_1(r)$, yielding
\begin{eqnarray}\label{case1t1}
    t_1(r)&=&-\,\frac{r\left[q_8'+2r\left(r t_4'+t_4\right)\right]+2 q_8}{2 r^2}-\frac{\left(q_8+2 r^2 t_4\right)}{4r}\left(\frac{\Psi_1'}{\Psi_{1}}+\frac{\Psi_2'}{\Psi_{2}}\right)\,.
\end{eqnarray}

Next, from $Y2^{\vartheta \vartheta r}=0$ we find that $q_7(r)$ reads
\begin{eqnarray}
   q_7(r)=\frac{1}{r^2\Psi_1 \Psi_2}\Big(c_4+\int \tilde{q}_7(r) dr \Big)\,,\label{case1q7}
\end{eqnarray}
where
\begin{eqnarray}
    \tilde{q}_7(r)&=&\frac{r}{4}\sqrt{\frac{\Psi_2}{\Psi_1}}\left(q_8+2 r^2 t_4\right) \Psi_1'^2-\frac{r}{2}\sqrt{\frac{\Psi_1}{\Psi_2}}\Big\{\Psi_1' \left[2 \Psi_2 \left(q_8'+2 r^2 t_4'\right)+2 r t_4 \left(r \Psi_2'+2 \Psi_2\right)+q_8 \Psi_2'\right]+\Psi_2 \left(q_8+2 r^2 t_4\right) \Psi_1''\Big\}\nonumber\\
    &&-\,\frac{\Psi_1}{4\Psi_2}\sqrt{\frac{\Psi_1}{\Psi_2}}\Big\{2 \Psi_2 \left[2 \Psi_2' \left(r q_8'+q_8+2 r^3 t_4'\right)+r q_8 \Psi_2''\right]+4 \Psi_2^2 \left[r\left(q_8''+2 r^2 t_4''+8 r t_4'\right)+2 q_8'\right]-r q_8 \Psi_2'^2\nonumber\\
    &&+\,2r t_4 \left[8 \Psi_2^2-r^2 \Psi_2'^2+2 r \Psi_2 \left(r \Psi_2''+4 \Psi_2'\right)\right]\Big\}\,.
\end{eqnarray}
Nevertheless, the Expressions~\eqref{case1t1} and~\eqref{case1q7} vanish the field strength tensor~\eqref{symmetric_contraction} for any value of $t_4(r)$ and $q_8(r)$, which trivially reduces the field equations to the Einstein equations and sets the Schwarzschild solution of GR as the only possible solution within this branch.

\subsubsection{Second main branch ($f_1\neq0$): case i) $f_2=-f_1/2$ }\label{f2f1/2}
If we assume that $f_2=-f_1/2$, from the equation~$Y2^{ttr}=0$ we find
\begin{equation}
t_1(r)=\frac{1}{4 r \Psi_1 \Psi_2}\Big[2\left(q_8'+2r^2t_4'\right)\Psi_1\Psi_2+\left(q_8+2r^2t_4\right)\bigl(\Psi_1\Psi_2'-\Psi_2 \Psi_1'\bigr)+4r\bigl(q_7\sqrt{\Psi_1} \Psi_2^{3/2}+t_4 \Psi_1 \Psi_2\bigr)\Big]\,.\label{casei}
\end{equation}
By using the above equation, and isolating $q_8''(r)$ from $Y2^{ttt}=0$ and $q_8'(r)$ from $Y2^{rrt}=0$ in terms of the rest of unknowns, one finds that the equation $Y2^{rrr}=0$ involves $q_8(r)=0$. This result solves directly the equation $Y2^{ttt}=0$, whereas $Y2^{rrt}=0$ is reduced to a more simple expression that allows one to isolate $t_4''(r)$. Then, the equation $Y2^{(tr)t}=0$ involves
\begin{eqnarray}
q_7(r)=-\,\frac{\sqrt{\Psi_1} }{\Psi_2^{3/2}}\left[2 r \Psi_2 t_4'+t_4\left(r \Psi_2'+2 \Psi_2\right)\right]\,.\label{case1q7B}
\end{eqnarray}
In this case, all the remaining connection field equations can be solved if the following differential equation is satisfied:
\begin{eqnarray}
t_4'(r)= t_4 \left(\frac{\Psi_1''}{\Psi_1'}-\frac{\Psi_1'}{2\Psi_1}-\frac{\Psi_2'}{2\Psi_2}-\frac{1}{r}\right)\,,
\end{eqnarray}
which sets the function $t_4(r)$ as
\begin{equation}
    t_4(r)=\frac{c_1\Psi_1'}{r\sqrt{\Psi_{1}\Psi_{2}}}\,,\label{t4is}
\end{equation}
where $c_1$ is a constant. Nevertheless, this value vanishes identically the corresponding expressions of the field strength tensors~\eqref{symmetric_curvature} and~\eqref{symmetric_contraction} of the nonmetricity field, which trivially sets the Schwarszchild solution of GR as the only solution of the field equations within this case.

\subsubsection{Second main branch ($f_1\neq0$): case ii)  $f_2\neq -f_1/2$}

In this section, we will assume $f_2\neq -f_1/2 \neq 0$. By isolating $t_1'(r)$ from $Y2^{\varphi\varphi r}=0$ and replacing its expression into the component $Y2^{\varphi\varphi t}$, we find
\begin{eqnarray}\label{formt1}
    t_1(r)&=&-\,\frac{1}{8 r^2 (f_1+2 f_2) \Psi_1^2 \Psi_2}\Big\{r \Psi_1 \Bigl[\Psi_1 \left[2 \Psi_2 \left(r \left(4 r (f_1+2 f_2) t_4'+f_1 q_8''\right)+4 f_2 q_8'\right)+3 f_1 r q_8' \Psi_2'\right]+f_1 r \Psi_2 q_8' \Psi_1'\nonumber\\
    &&+\,4 r (f_1+2 f_2) t_4 \left[r \Psi_2 \Psi_1'+\Psi_1 \left(r \Psi_2'+2 \Psi_2\right)\right]\Big]+q_8 \Big[r \Psi_1 \left[\Psi_2 \left(2 (f_1+2 f_2) \Psi_1'+f_1 r \Psi_1''\right)+f_1 r \Psi_1' \Psi_2'\right]\nonumber\\
    &&+\,\Psi_1^2 \left[r \left(f_1 r \Psi_2''-2 (f_1-2 f_2) \Psi_2'\right)+4 (f_1+4 f_2) \Psi_2\right]-f_1 r^2 \Psi_2 \Psi_1'^2\Big]\Big\}\,.\label{caseii}
    \end{eqnarray}
    
If one considers the expression~\eqref{formt1} to evaluate $t'_1(r)$ and $t_1(r)$ in the components $Y2^{\vartheta\vartheta t}$ and $Y2^{r t r}$ of the connection field equations, it is also possible to isolate $t_4''(r)$ and $q_8''(r)$, respectively. Then, by replacing their expressions into the component $Y2^{ttr}$, we find the following differential equation:
\begin{equation}\label{main_eq}
(f_1+4 f_2)\left[q_8+2 r (r t_4-1)\right]\Big\{2 r q_7 \Psi_2^{3/2}+\sqrt{\Psi_1} \left[2 \Psi_2 \left(q_8'+2 r^2 t_4'\right)+q_8 \Psi_2'\right]+2 r t_4 \sqrt{\Psi_1} \left(r \Psi_2'+2 \Psi_2\right)\Big\}=0\,.
\end{equation}

As can be seen, this equation can be solved in three different ways, giving rise to three subcases, which are analysed in detail below. 
%\parskip
\paragraph{Subcase 1. $q_8(r)=2r\left(1-rt_{4}(r)\right)$}%\parskip
Let us first explore the subcase 1. By solving $t_4'(r)$ from $Y2^{ttr}=0$ and replacing the resulting expression in~$Y2^{rrr}=0$, we find that
\begin{eqnarray}\label{exp_subcase1}
q_7(r)=\frac{\Psi_1 \left\{\Psi_2\left[2\left(2f_1+3f_2\right)\left(rt_4-1\right)-2f_{2}rt_{1}\right]-\left(f_{1}+f_{2}\right)r\Psi_2'\right\}-f_2 r \Psi_2 \Psi_1'}{f_1 r \sqrt{\Psi_1} \Psi_2^{3/2}}\,.
\end{eqnarray}
Now, by taking $Y2^{rrr}=0$ one finds only two possibilities, either $\Psi_1(r) = c_{1}r^{-\,\frac{4 f_2}{f_1+2 f_2}}$ or $\Psi_1(r) \neq c_{1}r^{-\,\frac{4 f_2}{f_1+2 f_2}}$, where $c_{1}$ is a constant. Nevertheless, it turns out that both of them lead to an inconsistent system of equations, which means the absence of solutions in the present subcase.

%\parskip
%%%%%%%%%%%%%%%%%%%%
\paragraph{Subcase 2. $q_7(r)=-\,\tfrac{\sqrt{\Psi_1} }{2 r \Psi_2^{3/2}}\left[2 \Psi_2 \left(q_8'+2 r^2 t_4'\right)+q_8 \Psi_2'+2 r t_4 \left(r \Psi_2'+2 \Psi_2\right)\right]$}
In this subcase, first it is straightforward to note that the connection field equations directly involve a vanishing symmetric curvature tensor if $f_{2}=0$, which means the absence of any dynamical solution with the traceless nonmetricity tensor under this condition. For $f_{2} \neq 0$, one can isolate $t_4''(r)$ from the connection equation $Y2^{\vartheta \vartheta r}=0$ and obtain the following expression for $t_{1}(r)$ from the equation $Y2^{ttr}=0$\,:
\begin{eqnarray}
t_1(r)&=&-\,\frac{1}{4 f_2 r^2 \Psi_1 \Psi_2}\Big\{
2f_{2}r\left[\Psi_1 \Psi_2 \left(q_8'+2 r^2 t_4'\right)+rt_{4}\left(r\Psi_2 \Psi_1'+r \Psi_1 \Psi_2'+2 \Psi_1 \Psi_2\right)\right]\nonumber\\
&&+\,q_8\bigl[\Psi_1 \left(2 (f_1+2 f_2) \Psi_2+f_2 r \Psi_2'\right)+f_2 r \Psi_2 \Psi_1'\bigr]
\Big\}\,.
\end{eqnarray}
From this expression, the antisymmetric part of the connection field equations in the present subcase is directly equivalent to $q_{8}(r)=0$, which in turn vanishes the symmetric part of the curvature tensor as well and sets the Schwarzschild geometry of GR as the only possible solution of the field equations.

\paragraph{Subcase 3. $f_2=-f_{1}/4$: exact Reissner-Nordstr\"{o}m-like solution with shear charges}\label{secsolution}
Let us now study the most important subcase, where it is possible to find a black hole solution with a dynamical traceless nonmetricity field. If we set $f_2=-f_{1}/4$ in the Expression~\eqref{formt1} and isolate again the functions $t_4''(r)$ and $q_8''(r)$ from the components $Y2^{\vartheta\vartheta t}$ and $Y2^{rtr}$ of the connection equations, respectively, we find that the combination $Y1^{t}{}_t-Y1^{r}{}_r = 0$ of the tetrad equations involves the reciprocity condition $\Psi_1(r)=\Psi_2(r)=\Psi(r)$ for the metric functions. In that case, the expression of $q_8''(r)$, previously obtained from the connection equation $Y2^{rtr}=0$, acquires the following simple form:
\begin{equation}
    q_{8}''=\frac{2r\left(\Psi q_{8}'+q_{8}\Psi'\right)-r^{2}\left(q_{8}\Psi''+2q_{8}'\Psi'\right)-2\Psi q_{8}}{r^{2}\Psi}\,,
\end{equation}
which can be directly solved, yielding
\begin{eqnarray}
q_8(r)=\frac{r\left(\kappa_{\rm sh}+c_{2}r\right)}{\Psi}\,,\label{q8}
\end{eqnarray}
where $\kappa_{\rm sh}$ and $c_2$ are integration constants describing the shear charges of the solution. Likewise, the expression of $t_4''(r)$ derived from $Y2^{\vartheta\vartheta t}=0$ reads
\begin{equation}
    t_{4}''=\frac{2\Psi\left(c_{2}r-\kappa_{\rm sh}\right)-2r^{3}\Psi\left(t_{4}\Psi''+2\Psi't_{4}'\right)-r^{2}\Psi\left(2t_{4}\Psi'+2\Psi' q_{7}+\Psi q_{7}'\right)+2r\Psi^{2}\left(q_{7}+2t_{4}\right)-r\Psi'\left(\kappa_{\rm sh}+c_{2}r\right)}{2r^{3}\Psi^{2}}\,,
\end{equation}
which allows us to set the value of $q_{7}(r)$ in terms of the functions $t_{4}(r)$ and $\Psi(r)$ as
\begin{eqnarray}
   q_7(r)=\frac{r^2}{\Psi^{2}(r)}\Big(c_4+\int \tilde{q}_7(r) dr \Big)\,,\label{case1q7important}
\end{eqnarray}
where
\begin{equation}\label{q7tilde}
    \tilde{q}_7(r)=\frac{1}{r^4}\Bigl[4rt_{4}\Psi^2+2\Psi\left(c_{2}r-\kappa_{\rm sh}\right)-r \Psi'\left(\kappa_{\rm sh}+c_{2}r\right)-2r^3\Psi\left(\Psi t_4''+2t_4'\Psi'+t_4\Psi''\right)-2r^{2}t_{4}\Psi\Psi'\Bigr]\,.
\end{equation}

In fact, this result vanishes the remaining connection field equations for any arbitrary value of $t_{4}(r)$, whereas the tetrad field equations are reduced to an independent differential equation involving the metric function alone
\begin{equation}
r^2\left(1-\Psi(r)\right)-r^3\Psi'(r) = -\,2f_{1}\kappa_{\rm sh}^{2}\,,
\end{equation}
which has as solution the Reissner-Nordstr\"{o}m metric depending on the square of the shear charge $\kappa_{\rm sh}$.

The final form of the solution of the field equations~\eqref{traceless_field_eq1} and~\eqref{traceless_field_eq2} is then given by
   \begin{eqnarray}
    t_1(r)&=&\frac{\kappa_{\rm sh}-2r^2\Psi(r) t_4'(r)-2r t_4(r) \left(r\Psi'(r)+\Psi(r)\right)}{2r\Psi(r)}\,,\quad q_8(r)=\frac{r\left(\kappa_{\rm sh}+c_{2}r\right)}{\Psi(r)}\,,\quad t_7(r)=0\,,\\
       q_7(r)&=&\frac{r^2}{\Psi(r)^2}\Big(c_4+\int \tilde{q}_7(r) dr \Big)\,, \\
   \Psi_1(r)&=&\Psi_2(r)=\Psi(r)=1-\frac{2m}{r}-\frac{2f_1 \kappa_{\rm sh}^2}{r^2}\,,\quad f_2=-\,\frac{1}{4}f_1\,,
   \end{eqnarray}
where $\tilde{q}_7(r)$ is defined in \eqref{q7tilde}, and the remaining components of torsion and nonmetricity are determined by the Expressions~\eqref{cond1},~\eqref{cond2a}-\eqref{cond2b},~\eqref{rel1_tors} and~\eqref{cond3a}-\eqref{cond3b}. Therefore, it turns out that the function $t_4(r)$ is totally free in the above solution. In fact, the appearance of arbitrary functions in the solutions of the field equations of MAG is a well-known result, which is also present in other models~\cite{Lenzen:1986hb,chen1994poincare,ho1997some}. In any case, in the next section we will show that such an arbitrariness can be fixed when considering the presence of a dynamical torsion field in the gravitational action.

On the other hand, considering the sign of the Lagrangian coefficient $f_1$, it is worthwhile to stress that the three irreducible parts of the traceless nonmetricity tensor are nonzero for general values of the function $t_{4}(r)$. Thereby, in order to avoid the kinetic energy of the vector $\Lambda_{\mu}$ to be unbounded from below, which would indicate the presence of a ghost instability in the solution, it is clear from the Expression~\eqref{model_expansion} that $f_{1} \leq 0$, which sets the geometry as the one described by the standard Reissner-Nordstr\"{o}m solution of GR, provided in this case by a traceless nonmetricity field rather than by an electromagnetic one.

\section{Reissner-Nordstr\"{o}m-like solutions with spin, dilation and shear charges}\label{sec:fullmodel}

The existence of Reissner-Nordstr\"{o}m-like solutions with spin and dilation charges $\kappa_{\rm s}$ and $\kappa_{\rm d}$, recently obtained in the realm of Weyl-Cartan geometry~\cite{Bahamonde:2020fnq}, together with our findings of Sec.~\ref{secsolution}, clearly suggests that both results should be described within a broad MAG model displaying the dynamics of the Weyl vector, as well as of the torsion and traceless nonmetricity tensors. Indeed, the models under consideration have in common that the geometrical corrections to GR are mediated by the torsion and nonmetricity fields alone, which in fact yield a nonvanishing metric curvature described by Einstein's model.

Following these lines, such a broad model must then introduce the field strength tensors of the torsion and nonmetricity fields in terms of all the corresponding deviations from the Bianchi identities of GR (including their contracted expressions), which are present in those particular models, namely
\begin{align}\label{curvbianchi}
\tilde{R}^{\lambda}\,_{[\mu \nu \rho]}&=\tilde{\nabla}_{[\mu}T^{\lambda}\,_{\rho\nu]}+T^{\sigma}\,_{[\mu\rho}\,T^{\lambda}\,_{\nu] \sigma}\,,\\
\tilde{R}_{[\mu \nu]}&=\frac{1}{2}\tilde{R}^\lambda\,_{\lambda\mu\nu}+\tilde{\nabla}_{[\mu}T^{\lambda}\,_{\nu]\lambda}+\frac{1}{2}\tilde{\nabla}_{\lambda}T^{\lambda}\,_{\mu\nu}-\frac{1}{2}T^{\lambda}\,_{\rho\lambda}T^{\rho}\,_{\mu\nu}\,,\label{riccibianchi}\\
\tilde{R}^{\lambda}\,_{\lambda\mu\nu}&=4\nabla_{[\nu}W_{\mu]}\,,
\end{align}
as well as the identities~\eqref{symmetric_curvature} and~\eqref{symmetric_contraction}. In particular, it is important to stress that the aforementioned expressions contain redundant information of the torsion and nonmetricity tensors, which must be taken into account in the construction of the gravitational action and the generalised solution. For instance, it is straightforward to note that the antisymmetric part of the Ricci tensor and the symmetric part of the curvature tensor include a homothetic component, which indeed constitutes the field strength tensor of the Weyl vector. Thereby, the final form of the Lagrangian must compensate all these redundant terms, which involves a particular choice of its coefficients and the presence of cross-products among the different parts of the curvature tensor.

The result accounting for these features depends then on three independent coefficients, associated with the three different dynamical parts of the affine connection, and reads
\begin{align}\label{full_action}
S = &\,\frac{1}{64\pi}\int
\Bigl[
-\,4R-6d_{1}\tilde{R}_{\lambda\left[\rho\mu\nu\right]}\tilde{R}^{\lambda\left[\rho\mu\nu\right]}-9d_{1}\tilde{R}_{\lambda\left[\rho\mu\nu\right]}\tilde{R}^{\mu\left[\lambda\nu\rho\right]}+2d_{1}\left(\tilde{R}_{[\mu\nu]}+\hat{R}_{[\mu\nu]}\right)\left(\tilde{R}^{[\mu\nu]}+\hat{R}^{[\mu\nu]}\right)
\Bigr.
\nonumber\\
\Bigl.
&+18d_{1}\tilde{R}_{\lambda\left[\rho\mu\nu\right]}\tilde{R}^{(\lambda\rho)\mu\nu}-3d_{1}\tilde{R}_{(\lambda\rho)\mu\nu}\tilde{R}^{(\lambda\rho)\mu\nu}+6d_{1}\tilde{R}_{(\lambda\rho)\mu\nu}\tilde{R}^{(\lambda\mu)\rho\nu}+2\left(2e_{1}-f_{1}\right)\tilde{R}^{\lambda}\,_{\lambda\mu\nu}\tilde{R}^{\rho}\,_{\rho}\,^{\mu\nu}
\Bigr.
\nonumber\\
\Bigl.
&+8f_{1}\tilde{R}_{(\lambda\rho)\mu\nu}\tilde{R}^{(\lambda\rho)\mu\nu}-2f_{1}\left(\tilde{R}_{(\mu\nu)}-\hat{R}_{(\mu\nu)}\right)\left(\tilde{R}^{(\mu\nu)}-\hat{R}^{(\mu\nu)}\right)+3\left(1-2a_{2}\right)T_{[\lambda\mu\nu]}T^{[\lambda\mu\nu]}
\Bigr]d^4x\sqrt{-g}\,,
\end{align}
which gives rise to the following independent field equations:
\begin{eqnarray}
X1_{\mu}\,^{\nu} &=& 0\,,\label{full_tetradeq}\\
X2^{\lambda\mu\nu} &=& 0,\label{full_connectioneq}
\end{eqnarray}
where $X1_{\mu}\,^{\nu}$ and $X2^{\lambda\mu\nu}$ are tensor quantities defined in~\eqref{Tensor_tetradeq}-\eqref{Tensor_connectioneq}.

Thereby, the action~\eqref{full_action} describes the same model considered in~\cite{Bahamonde:2020fnq} when the traceless nonmetricity tensor vanishes. In order to solve its field equations, one can then combine the solution presented in~\cite{Bahamonde:2020fnq}, in the presence of dynamical torsion and Weyl vector, with the one obtained in Sec.~\ref{secsolution}. This means we can consider the reciprocity condition for the metric functions and a general form for all the torsion and nonmetricity components as a combination $q_{i}(r)=q_{\textrm{a},i}(r)+q_{\textrm{b},i}(r)$ and $t_{i}(r)=t_{\textrm{a},i}(r)+t_{\textrm{b},i}(r)$, where ``$\textrm{a}$" refers to the solutions originally obtained in~\cite{Bahamonde:2020fnq} and ``$\textrm{b}$" to the solutions obtained in Sec.~\ref{secsolution}. Applying this procedure without fixing the expression of $q_{\textrm{b},7}(r)$ and of the arbitrary function $t_{\textrm{b},4}(r)$, the torsion and nonmetricity functions read
\begin{eqnarray}
t_1(r)&=&\frac{\Psi'(r)}{2\Psi(r)}+\frac{wr}{\Psi(r)}+\frac{\kappa_{\rm d}+\kappa_{\rm sh}}{2r\Psi(r)}-r t'_{\textrm{b},4}(r)-\frac{t_{\textrm{b},4}(r) \left(r\Psi'(r)+\Psi(r)\right)}{\Psi(r)}\,,\label{final_form1}\\
t_2(r)&=&t_1(r)\Psi(r)\,,\quad t_3(r)=-\,t_4(r)\Psi(r)\,,\\
t_4(r)&=&-\,\frac{1}{2r}-\frac{wr}{2\Psi(r)}-\frac{\kappa_d}{2r\Psi(r)}+t_{\textrm{b},4}(r)\,,\quad \\
t_5(r)&=&-\,t_6(r)\Psi(r)\,,\quad t_6(r)=\frac{\kappa_{\rm s}}{r\Psi(r)}\,,\quad t_7(r)=t_8(r)=0\,,\\
q_1(r)&=&\frac{\Psi(r)}{r}\Big\{\kappa_{\rm d}-3\kappa_{\rm sh}-2 c_2r+rq_{\textrm{b},7}(r)\Psi(r)+2r\left[r\Psi(r) t'_{\textrm{b},4}(r)+t_{\textrm{b},4}(r) \left(r\Psi'(r)+\Psi(r)\right)\right]\Bigr\}\,,\\
q_2(r)&=&\frac{q_1(r)}{\Psi^{2}(r)}+\frac{2\left(\kappa_{\rm sh}-\kappa_{\rm d}+c_{2}r\right)}{r\Psi(r)}\,,\quad 
q_3(r)=-\,\frac{q_1(r)}{\Psi(r)}+\frac{\kappa_{\rm d}-\kappa_{\rm sh}-c_{2}r}{r}\,,\\q_4(r)&=&-\,r\left(\kappa_{\rm d}+\kappa_{\rm sh}+c_{2}r\right)\,,\quad q_5(r)=\frac{\kappa_{\rm sh}-\kappa_{\rm d}+c_2 r-rq_{\textrm{b},7}(r) \Psi(r)}{r}\,,\\
q_6(r)&=&-\,\frac{q_5(r)}{\Psi^{2}(r)}-\frac{2q_{\textrm{b},7}(r)}{\Psi(r)}\,,\quad q_7(r)=q_{\textrm{b},7}(r)\,,\\
q_8(r)&=&-\,\frac{q_4(r)}{\Psi(r)}\,,\quad 
 q_9(r)=-\,r\left(\kappa_{\rm sh}+c_2 r+2r t_{\textrm{b},4}(r)\Psi(r)\right)\,,\quad q_{10}(r)=-\,\frac{q_9(r)}{\Psi(r)}\,,\quad q_{11}(r)=q_{12}(r)=0\,,\label{final_form9}
\end{eqnarray}
with $w=(1-2a_{2})/d_{1}$. These expressions turn out to satisfy the antisymmetric and trace parts of the connection field equations~\eqref{full_connectioneq}, whereas the symmetric traceless part is reduced to the following independent equations:
\begin{align}
    0=&f_{1}\bigl[
    -2r^2\Psi q_{\textrm{b},7}\Psi'-2r^2\Psi t_{\textrm{b},4}\Psi'-4wr^{4}t_{\textrm{b},4}\Psi'-4\Psi \kappa_{\rm sh}+2r\Psi^{2}q_{\textrm{b},7}+4r\Psi^{2}t_{\textrm{b},4}-4wr^{4}\Psi t_{\textrm{b},4}'+2r\Psi' \kappa_{\rm sh}\nonumber\\
    &+c_{2}r^{2}\Psi'-6r^{3}\Psi\Psi't_{\textrm{b},4}'+4w\kappa_{\rm sh}r^{2}-2r^{3}\Psi'^{2}t_{\textrm{b},4}+2wc_{2}r^3-r^2\Psi^{2}q_{\textrm{b},7}'-2r^3\Psi^{2}t_{\textrm{b},4}''-2r^3\Psi t_{\textrm{b},4}\Psi''
    \bigr]\,,\label{finaleq1}\\
    0=&\,\kappa_{\rm s}\left\{2r t_{\textrm{b},4} \left[r\left(d_{1}+8f_{1}\right)\Psi'+2d_{1}\Psi\right]-\left(d_{1}+8f_{1}\right) (\kappa_{\rm sh}-2r^2\Psi t_{\rm b,4}')\right\}\,\label{finaleq2}.
\end{align}

Hence, from Eq.~\eqref{finaleq1} it is possible to determine the form of $q_{\textrm{b},7}(r)$ as an integral expression
\begin{eqnarray}\label{formq7}
   q_{\textrm{b},7}(r)=\frac{r^2}{\Psi^{2}(r)}\Big(c_4+\int\tilde{q}_{\textrm{b},7}(r) dr \Big)\,,\label{case1q7important}
\end{eqnarray}
where
\begin{align}
    \tilde{q}_{\textrm{b},7}(r)=&\,\frac{1}{r^4}\Bigl\{r\Psi'\left[2\kappa_{\rm sh}+c_{2}r-2r\left(\Psi t_{\textrm{b},4}+rt_{\textrm{b},4}\Psi'+3r\Psi t_{\textrm{b},4}'\right)\right]-4\Psi\left(\kappa_{\rm sh}-r\Psi t_{\textrm{b},4}\right)\nonumber\\
    &-2r^{3}\Psi\left(\Psi t_{\textrm{b},4}''+t_{\textrm{b},4}\Psi''\right)+2wr^{2}\left[2\kappa_{\rm sh}+c_{2}r-2r^{2}\left(t_{\textrm{b},4}\Psi'+\Psi t_{\textrm{b},4}'\right)\right]\Bigr\}\,.
\end{align}

On the other hand, from Eq.~\eqref{finaleq2} it is clear that if $\kappa_{\rm s}=0$ the form of $t_{\textrm{b},4}(r)$ is left arbitrary, which means that the presence of dynamical torsion constrains the form of the traceless nonmetricity tensor in the field equations. In any case, the resulting expression of $t_{\textrm{b},4}(r)$ can also differ, depending on the values of the Lagrangian coefficients $d_{1}$ and $f_{1}$. In particular, if $\kappa_{\rm s} \neq 0$ one can distinguish three different solutions for $t_{\textrm{b},4}(r)$ in Eq.~\eqref{finaleq2}, each one arising from the cases $d_1=8f_1$, $d_1=-\,8f_1$ and $d_1\neq \pm \, 8f_1$.

Thus, the components of the torsion and nonmetricity tensors of the complete solution have the form~\eqref{final_form1}-\eqref{final_form9}, with the functions $t_{\textrm{b},4}(r)$ and $\tilde{q}_{\textrm{b},7}(r)$ presented in Table~\ref{tab:final_functions} for each of the aforementioned cases.
\begin{table}[H]
    \centering
  \bgroup
\def\arraystretch{3}  \begin{tabular}{|c|c|c|}
    \hline
        Theory & $t_{\textrm{b},4}(r)$ & $\tilde{q}_{\textrm{b},7}(r)$ \\ \hline    \multirow{3}{*}{$d_1\neq \pm 8f_1$}  & \multirow{3}{*}{$\displaystyle\frac{1}{2\Psi(r)}\left[
        \frac{(d_1+8 f_1)\,\kappa_{\rm sh}}{(d_1-8 f_1)\,r}+2 c_3 r^{-\,\frac{2 d_1}{d_1+8 f_1}}
        \right]$} & $\displaystyle\frac{2 t_{\textrm{b},4}(r) \Psi(r)}{r^3 (d_1+8f_1)^2} \Big\{r\left(d_1+8 f_1\right) \left[\left(d_1-8f_1\right) \Psi'(r)+4 d_1 wr\right]$ \\ 
       & & \hspace{1.5cm}$-\,4\left(d_1^2-4 d_1 f_1-32 f_1^2\right)\Psi(r)\Big\}$  \\ 
       & &  \hspace{-1cm}$\displaystyle+\,\frac{\left(\kappa_{\rm sh}+c_{2}r\right)\left(2wr+\Psi'(r)\right)}{r^3}-\displaystyle\frac{16 f_1 \kappa_{\rm sh} \Psi(r)}{\left(d_1+8 f_1\right)r^4}$\\[2ex]\hline
       $d_1=8f_1$  & $\displaystyle\frac{2 c_3+\kappa_{\rm sh}\log(r)}{2r\Psi(r)}$ & $\displaystyle\frac{r\left(\kappa_{\rm sh}+c_{2}r\right)\left(2wr+\Psi'(r)\right)-\Psi(r)\left(\kappa_{\rm sh}-4wr^{3}t_{\textrm{b},4}(r)\right)}{r^4}$\\[2ex] \hline
       $d_1=-\,8f_1$  & 0 & $\displaystyle\frac{r\left(2\kappa_{\rm sh}+c_{2}r\right)\left(2wr+\Psi'(r)\right)-4\kappa_{\rm sh}\Psi(r)}{r^4}$\\[2ex] \hline
    \end{tabular}\egroup
    \caption{Functions $t_{\textrm{b},4}(r)$ and $\tilde{q}_{\textrm{b},7}(r)$ of the complete solution.}
    \label{tab:final_functions}
\end{table}

Finally, the resulting tetrad field equations~\eqref{full_tetradeq} allow the metric function $\Psi(r)$ to be determined, for any value of the Lagrangian coefficients, from the following differential expression:
\begin{equation}
r^2\left(1-\Psi(r)\right)-r^3\Psi'(r) = d_{1}\kappa_{\rm s}^{2}-4e_{1}\kappa_{\rm d}^{2}-2f_{1}\kappa_{\rm sh}^{2}\,,
\end{equation}
which provides a Reissner-Nordstr\"{o}m-like geometry with spin, dilation and shear charges
\begin{eqnarray}
    \Psi(r)&=&1-\frac{2 m}{r}+\frac{d_1 \kappa_{\rm s}^2-4 e_1 \kappa_{\rm d}^2-2f_1 \kappa_{\rm sh}^2}{r^2}\,.
\end{eqnarray}
Note that one can easily integrate Eq.~\eqref{formq7} for the three sets of solutions by replacing the above form of $\Psi(r)$.

Being the components of the torsion and nonmetricity tensors already known, it is then straightforward to compute their irreducible parts and the specific contribution of these parts to the field strength tensors of the model. In particular, it turns out that there are certain parts in the torsion sector whose differential contribution to the field strength tensors vanishes and accordingly they do not generate any kinetic term in the gravitational action. Thereby, we can denote these quantities by a circle on top and the ones providing kinetic terms in the action by a bar, which gives rise to the following form for the irreducible parts of the torsion tensor of the solution:
\begin{eqnarray}
T_{\mu}&=&\mathring{T}_{\mu}\,,\\
S_{\mu}&=&\bar{S}_{\mu}\,,\\
t_{\lambda\mu\nu}&=&\mathring{t}_{\lambda\mu\nu}+\bar{t}_{\lambda\mu\nu}\,,
\end{eqnarray}
Thus, the expressions of the field strength tensors in terms of the corresponding irreducible parts of the torsion and nonmetricity fields of the solution read\footnote{See the list of expressions of the irreducible parts of the solution in Appendix~\ref{sec:AppIrrPartsSol}.}:
\begin{eqnarray}
    \tilde{R}^{(\lambda \rho)}{}_{\mu \nu}&=&-\,\frac{1}{2}g^{\lambda \rho}F^{(W)}_{\mu\nu}+{\nearrow\!\!\!\!\!\!\!\tilde{R}}^{(\lambda \rho)}{}_{\mu \nu}\,,\\
    \tilde{R}_{(\mu\nu)}-\hat{R}_{(\mu\nu)}&=&2\nabla_{(\mu}\Lambda_{\nu)}-\frac{1}{2}\,g_{\mu\nu}\nabla_{\lambda}\Lambda^{\lambda}+\frac{1}{6}\,\varepsilon^{\sigma}\,_{\lambda\rho(\mu|}\nabla_{\sigma}\Omega_{|\nu)}\,^{\lambda\rho}-\nabla_{\lambda}q^{\lambda}\,_{\mu\nu}-\frac{4}{3}\mathring{T}_{1(\mu}\Lambda_{\nu)}-\frac{5}{2}\mathring{t}_{1(\mu\nu)\lambda}\Lambda^{\lambda}\nonumber\\
    &&+\,\frac{1}{18}\varepsilon_{\sigma\omega\lambda(\mu}\Omega_{\nu)}{}^{\sigma\omega}\mathring{T}_{1}^{\lambda}+\frac{5}{6}\varepsilon_{\sigma\omega\lambda\rho}\Omega_{(\mu}{}^{\sigma\omega}\mathring{t}_{1}^{\lambda\rho}{}_{\nu)}-\frac{4}{3}q_{\mu\nu\lambda}\mathring{T}_{1}^{\lambda}-q_{\lambda\rho(\mu}\mathring{t}_{1}^{\lambda\rho}{}_{\nu)}\,,\\
    \tilde{R}^{\sigma}\,_{\sigma\mu\nu}&=&-\,2F^{(W)}_{\mu\nu}\,,\\
    \tilde{R}^{\lambda}\,_{[\mu\nu\rho]}&=&\frac{1}{4}\tilde{R}^{\sigma}\,_{\sigma[\mu\nu}\delta_{\rho]}\,^{\lambda}-\frac{1}{2}\check{R}_{[\mu\nu}\delta_{\rho]}\,^{\lambda}+\bar{R}^{\lambda}\,_{[\mu \nu \rho]}\,,\\
    \tilde{R}_{[\mu\nu]}&=&\frac{1}{4}\tilde{R}^{\sigma}\,_{\sigma\mu\nu}+\frac{1}{2}\check{R}_{\mu\nu}+\bar{R}_{[\mu\nu]}\,,\\
    \hat{R}_{[\mu\nu]}&=&-\,\frac{1}{4}\tilde{R}^{\sigma}\,_{\sigma\mu\nu}-\frac{1}{2}\check{R}_{\mu\nu}+\bar{R}_{[\mu\nu]}\,,
\end{eqnarray}
where
\begin{eqnarray}
    {\nearrow\!\!\!\!\!\!\!\tilde{R}}^{(\lambda \rho)}{}_{\mu \nu}&=&\frac{1}{8}\,g^{\lambda \rho}F^{(\Lambda)}_{\mu\nu}-\nabla_{[\mu}\Lambda^{(\lambda}\delta_{\nu]}^{\rho)} + \frac{1}{3}\,\varepsilon_{\sigma\omega[\mu}\,^{(\lambda}\nabla_{\nu]}\Omega^{\rho)\sigma\omega}-\nabla_{[\mu}q^{\lambda \rho}\,_{\nu]}-\frac {2} {3}\delta^{(\lambda} _ {[\mu}\Lambda^{\rho)} \mathring{T}_{1\nu]}-\delta^{(\lambda} _ {[\mu} \mathring{t}_{1}^{\rho)} {} _ {\nu] \sigma}\Lambda^\sigma+\frac{1}{2}\Lambda^{(\lambda}\mathring{t}_{1}^{\rho)} {} _ {\mu\nu}\nonumber\\
    &&-\,\frac{1}{6}\varepsilon_ {\sigma\omega\mu\nu} \mathring{T}_{1}^{(\lambda}\Omega^{\rho)\sigma\omega}-\frac{1}{9}\varepsilon_ {\sigma\omega\gamma[\mu}\delta^{(\lambda} {} _ {\nu]}\Omega^{\rho)\sigma\omega} \mathring{T}_{1}^{\gamma}+\,\frac{2}{3}\mathring{t}_{1}^{\gamma(\lambda} {} _ {[\mu}\varepsilon_ {\nu] \gamma\sigma\omega}\Omega^{\rho)\sigma\omega}-\frac{1}{2}\mathring{t}_ {1\gamma}{}^{(\lambda}{} _ {[\mu}\varepsilon^{\rho)} {} _ {\nu]\sigma\omega}\Omega^{\gamma\sigma\omega}\nonumber\\
    &&-\,\frac{2}{3}\delta^{(\lambda} _ {[\mu} q^{\rho)} {} _ {\nu] \sigma} \mathring{T}_{1}^\sigma-2q^{(\lambda} {} _ {\sigma[\mu} \mathring{t}_ {1\nu]} {}^{\rho) \sigma}-\frac{1}{12}\varepsilon_{\mu\nu\sigma}{}^{(\lambda}\Lambda^{\rho)} \bar{S}^\sigma+\frac{1}{2}\bar{t}^{(\lambda} {} _ {\mu\nu}\Lambda^{\rho)}-\frac{1}{3}\bar{t}^{\gamma(\lambda} {} _ {[\mu}\varepsilon_ {\nu] \gamma\sigma\omega}\Omega^{\rho)\sigma\omega}\nonumber\\
    &&-\frac{1}{6}\varepsilon_{\sigma\omega[\mu}{}^{(\lambda}q^{\rho)}{}_{\nu]}{}^{\sigma}\bar{S}^{\omega}-2q^{(\lambda} {} _ {\sigma[\mu} \bar{t}_ {\nu]} {}^{\rho)\sigma}\,,\\
    \check{R}_{\mu\nu}&=&-\,\frac{3}{4}F^{(\Lambda)}_{\mu\nu}+\frac{1}{6}\left(\varepsilon_{\lambda\rho\sigma[\mu}\nabla_{\nu]}\Omega^{\lambda\rho\sigma}+\varepsilon_{\lambda\rho\mu\nu}\nabla_{\sigma}\Omega^{\sigma\lambda\rho}-\varepsilon^{\sigma}\,_{\lambda\rho[\mu|}\nabla_{\sigma}\Omega_{|\nu]}\,^{\lambda\rho}\right)\,,\\
    \bar{R}^{\lambda}\,_{[\mu \nu \rho]}&=&\frac{1}{6}\varepsilon^{\lambda}\,_{\sigma[\rho\nu}\nabla_{\mu]}\bar{S}^{\sigma}+\nabla_{[\mu}\bar{t}^{\lambda}\,_{\rho\nu]}-\frac{1}{18}\varepsilon_{\sigma\mu\nu\rho}\mathring{T}_{1}^{\lambda}\bar{S}^{\sigma}+\frac{1}{4}\varepsilon^{\lambda}\,_{\omega\sigma[\rho}\mathring{t}_{1}^{\sigma}\,_{\mu\nu]}\bar{S}^{\omega}\nonumber\\
    &&+\,\frac{1}{24}\varepsilon_{\mu\nu\rho\sigma}\bar{S}^{\sigma}\Lambda^{\lambda}-\frac{1}{4}\varepsilon^{\sigma\omega\lambda\tau}\Omega_ {\gamma\sigma\omega}\bar{t}^\gamma {} _ {[\mu \nu} g_ {\rho]\tau}+\frac{1}{4}\varepsilon_ {[\mu\nu} {}^{\sigma\omega} q^\lambda {} _ {\rho]\sigma}\bar{S}_\omega\,,\\
    \bar{R}_{[\mu\nu]}&=&\frac{1}{12}\varepsilon^{\lambda}\,_{\sigma\mu\nu}\nabla_{\lambda}\bar{S}^{\sigma}+\frac{1}{2}\nabla_{\lambda}\bar{t}^{\lambda}\,_{\mu\nu}\,.
\end{eqnarray}

Furthermore, the effective gravitational action of the model expressed in terms of these quantities acquires the following form:
\begin{eqnarray}
    S &=& \frac{1}{64 \pi}\int d^4x \sqrt{-g}
    \Bigl[
    -\,4R-6d_{1}\bar{R}_{\lambda\left[\rho\mu\nu\right]}\bar{R}^{\lambda\left[\rho\mu\nu\right]}-9d_{1}\bar{R}_{\lambda\left[\rho\mu\nu\right]}\bar{R}^{\mu\left[\lambda\nu\rho\right]}+8\,d_{1}\bar{R}_{\left[\mu\nu\right]}\bar{R}^{\left[\mu\nu\right]}+4e_{1}\tilde{R}^{\lambda}\,_{\lambda\mu\nu}\tilde{R}^{\rho}\,_{\rho}\,^{\mu\nu}
    \Bigr.
    \nonumber\\& &
    \left.
    \Bigl.
    +\,8f_1{\nearrow\!\!\!\!\!\!\!\tilde{R}}_{(\lambda\rho)\mu\nu}{\nearrow\!\!\!\!\!\!\!\tilde{R}}^{(\lambda\rho)\mu\nu}-2 f_{1}\left(\tilde{R}_{(\mu\nu)}-\hat{R}_{(\mu\nu)}\right)\left(\tilde{R}^{(\mu\nu)}-\hat{R}^{(\mu\nu)}\right)+18 d_1 \bar{R}_{\mu [\lambda \rho \nu]}{\nearrow\!\!\!\!\!\!\!\tilde{R}}^{(\mu \nu)\lambda \rho}-6d_{1}\bar{R}_{[\mu\nu]}\check{R}^{\mu\nu}\right.\nonumber\\
    &&-\,3d_1{\nearrow\!\!\!\!\!\!\!\tilde{R}}_{(\lambda\rho)\mu\nu}{\nearrow\!\!\!\!\!\!\!\tilde{R}}^{(\lambda\rho)\mu\nu}+6 d_1 {\nearrow\!\!\!\!\!\!\!\tilde{R}}_{(\lambda \rho)\mu\nu} {\nearrow\!\!\!\!\!\!\!\tilde{R}}^{(\lambda\mu)\rho\nu}+\frac{9}{2}d_{1}\check{R}_{\mu\nu}\check{R}^{\mu\nu}+3\left(1-2a_{2}\right)T_{\left[\lambda\mu\nu\right]}T^{\left[\lambda\mu\nu\right]}
    \Bigr]\,,
\end{eqnarray}
which clearly displays interaction terms between the dynamical torsion and traceless nonmetricity tensors, preventing the degeneracy obtained in the results of the previous section in the absence of this interaction.

On the other hand, the solution can also be trivially generalised to include the cosmological constant and Coulomb electromagnetic fields with electric and magnetic charges $q_{\rm e}$ and $q_{\rm m}$, which are decoupled from torsion and nonmetricity under the assumption of the minimal coupling principle. This natural extension is then simply achieved by replacing the metric function by the one described by the Reissner-Nordstr\"{o}m-de Sitter-like geometry
\begin{equation}
\Psi(r)=1-\frac{2m}{r}+\frac{d_{1}\kappa^{2}_{\rm s}-4e_{1}\kappa^{2}_{\rm d}-2f_1 \kappa_{\rm sh}^2+q^{2}_{\rm e}+q^{2}_{\rm m}}{r^2}+\frac{\Lambda}{3}r^{2}\,,
\end{equation}
which turns out to represent the broadest family of static and spherically symmetric black hole solutions obtained in MAG so far.

\section{Conclusions}\label{sec:conclusions}

In this work, we have considered a MAG model which displays the dynamics of the traceless nonmetricity tensor as a minimal deviation from GR. As a direct consequence, the Birkhoff's theorem of GR is no longer held, which allows us to find, under certain consistency constraints, a new Reissner-Nordstr\"{o}m-like black hole solution endowed with shear charges. From a physical point of view, this quantity can be considered as a measure for the volume-preserving deformation of matter interacting through quadrupole moments~\cite{hehl1991spacetime}, which for the case of hadronic matter reproduces the parent and daughter Regge trajectories with angular momentum transitions of $|\Delta J|=2$~\cite{Dothan:1965aa,Neeman:1987nsi}.

The form of both the model and the solution resembles the previous findings obtained in~\cite{Bahamonde:2020fnq}, in the context of Weyl-Cartan geometry, which suggests the existence of a broader MAG model describing all these results. Accordingly, we introduce a gravitational action with second order corrections provided by the torsion and nonmetricity fields analysed so far, and obtain a generalised Reissner-Nordstr\"{o}m-like black hole solution with spin, dilation and shear charges. Thereby, the resulting geometry displays a full dynamical correspondence between the metric, the coframe and the three main independent post-Riemannian parts present in the connection, namely the torsion tensor, as well as the trace and traceless componentes of the nonmetricity tensor, which constitutes the broadest family of static and spherically symmetric black hole solutions obtained in MAG so far.

The aforementioned findings achieved by including the traceless part of the nonmetricity tensor into the geometrical framework open the door not only to further phenomenological analyses on this sector of the theory (see~\cite{Yagi:2016jml,Bambi:2015kza,Berti:2018cxi,Barack:2018yly,Bahamonde:2021akc,EventHorizonTelescope:2021dqv,Kramer:2021jcw,Daas:2022iid,EventHorizonTelescope:2022xqj,Pantig:2022ely} for a recent list of observational constraints on extended theories of gravity), but also to new studies concerning the stability of the tensor modes displayed by the action~\eqref{full_action}. Indeed, whereas the stability of the scalar and vector modes of the torsion and nonmetricity tensors has been analysed in detail for quadratic MAG models~\cite{Jimenez:2019qjc,Percacci:2020ddy,BeltranJimenez:2020sqf,Baldazzi:2021kaf,Jimenez-Cano:2022sds}, a thorough study concerning the behaviour of the tensor parts is yet to be performed.

Likewise, the recent formulation of the Hamiltonian approach to black hole entropy in the framework of Poincar\'{e} gauge theory~\cite{Blagojevic:2019gsd}, has shown to provide a consistent thermodynamics in the torsion sector of the solution~\cite{Blagojevic:2021pqp} (see~\cite{Blagojevic:2019bqg,Blagojevic:2020edq,Blagojevic:2020ymf,Blagojevic:2021mli,Blagojevic:2022etm,Cvetkovic:2022qpt} for further analyses considering other black hole solutions with and without torsion). In this sense, our results also highlight the need to extend the Hamiltonian formulation developed in~\cite{Blagojevic:2019gsd} towards general metric-affine geometries, where the nonmetricity tensor is another relevant quantity, in order to describe the entropy of the full solution and analyse the contribution of the spin, dilation and shear charges to the first law of black hole thermodynamics.

From a quantum field theory perspective, the spectrum of the Hawking radiation in the presence of both dynamical torsion and nonmetricity fields is yet to be determined. In any case, a general splitting of the energy levels of those matter currents coupled to torsion and nonmetricity will potentially affect this spectrum as well as the efficiency of the emission process, in comparison with the standard case of GR. Of particular interest is the performance of a perturbative analysis on the vacuum fluctuations of the energy-momentum tensor of a quantum field coupled to the torsion and nonmetricity tensors of the solution, in order to examine the rate of dissipation induced on its event horizon, which would also involve further corrections with respect to the framework of GR~\cite{Campos:1998xq,hu1999notes} (see also~\cite{Hensh:2022pmh} for a recent analysis on tidal heating in Einstein-Cartan theory with torsion).

Finally, it is also worthwhile to mention other interesting prospects of research, such as the extension of these results in the presence of odd parity modes of torsion and nonmetricity, and/or in axisymmetric space-times. On the other hand, the cosmological implications and the contribution to the spectrum of gravitational waves of the proposed model turn out to be other important questions to be investigated. Further studies concerning these lines are currently underway.

\bigskip
\bigskip
\noindent
\section*{Acknowledgements}

The authors would like to thank Kostas Pallikaris for his contribution during the initial stages of this work. S.B. is supported by JSPS Postdoctoral Fellowships for Research in Japan and KAKENHI Grant-in-Aid for
Scientific Research No. JP21F21789. J.G.V. is supported by the European Regional Development Fund through the Center of Excellence TK133 ``The Dark Side of the
Universe". J.C. is grateful to the warm hospitality of the University of Tartu, where this work was completed and written.

\newpage

\appendix
\section{Field equations}\label{sec:AppFieldEqs}

The variation of the action (\ref{TracelessLagrangian}) provides the following tensor quantities depending on the curvature, torsion and nonmetricity tensors:
\begin{eqnarray}
Y1_{\mu}\,^{\nu} &=& G_{\mu}\,^{\nu}+4f_{1}\tilde{R}_{(\lambda\rho)\mu\sigma}\tilde{R}^{(\lambda\rho)\sigma\nu}+2f_{2}\left(\tilde{R}_{\lambda\mu}-\hat{R}_{\lambda\mu}\right)\left(\hat{R}^{(\lambda\nu)}-\tilde{R}^{(\lambda\nu)}\right)
\nonumber\\
&&+\,2f_{2}\left(\tilde{R}^{\nu}\,_{\lambda\rho\mu}-\tilde{R}_{\lambda}\,^{\nu}\,_{\mu\rho}\right)\left(\tilde{R}^{(\lambda\rho)}-\hat{R}^{(\lambda\rho)}\right)+8\pi\tilde{\mathcal{L}}_{Q}\,\delta_{\mu}\,^{\nu}\,,\label{tetradeqcase1}
\end{eqnarray}
\begin{eqnarray}
Y2^{\lambda\mu\nu}&=&-\,f_{2}\Bigl[
g^{\lambda\nu}\left(\nabla_{\rho}\tilde{R}^{(\mu\rho)}-\nabla_{\rho}\hat{R}^{(\mu\rho)}\right)+g^{\lambda\rho}\left(\nabla_{\rho}\hat{R}^{(\mu\nu)}-\nabla_{\rho}\tilde{R}^{(\mu\nu)}\right)+g^{\mu\nu}\left(\nabla_{\rho}\tilde{R}^{(\lambda\rho)}-\nabla_{\rho}\hat{R}^{(\lambda\rho)}\right)
\Bigr.\,
\nonumber\\
\Bigl.
&&+\,g^{\mu\rho}\left(\nabla_{\rho}\hat{R}^{(\lambda\nu)}-\nabla_{\rho}\tilde{R}^{(\lambda\nu)}\right)+N^{\mu\nu}\,_{\rho}\left(\tilde{R}^{(\lambda\rho)}-\hat{R}^{(\lambda\rho)}\right)+N^{\nu\lambda}\,_{\rho}\left(\hat{R}^{(\mu\rho)}-\tilde{R}^{(\mu\rho)}\right)+N^{\rho\lambda}\,_{\rho}\left(\tilde{R}^{(\mu\nu)}-\hat{R}^{(\mu\nu)}\right)
\Bigr.\,
\nonumber\\
\Bigl.
&&+\,\Bigl(g^{\lambda\nu}N^{\mu}\,_{\sigma\rho}-g^{\mu\nu}N_{\sigma}\,^{\lambda}\,_{\rho}\Bigr)\left(\tilde{R}^{(\sigma\rho)}-\hat{R}^{(\sigma\rho)}\right)+\Bigl(N_{\rho}\,^{\lambda\mu}-N^{\mu}\,_{\rho}\,^{\lambda}\Bigr)\left(\tilde{R}^{(\nu\rho)}-\hat{R}^{(\nu\rho)}\right)+N^{\mu\rho}\,_{\rho}\left(\hat{R}^{(\lambda\nu)}-\tilde{R}^{(\lambda\nu)}\right)\Bigr]
\nonumber\\
&&+\,2f_{1}\left(\nabla_{\rho}\tilde{R}^{(\mu\lambda)\rho\nu}+N^{\mu}\,_{\sigma\rho}\tilde{R}^{(\lambda\sigma)\rho\nu}-N_{\sigma}\,^{\lambda}\,_{\rho}\tilde{R}^{(\mu\sigma)\rho\nu}\right)\,,\label{connectioneqcase1}
\end{eqnarray}
in such a way that the correspondence with the canonical energy-momentum and hypermomentum tensors gives rise to the field equations:
\begin{eqnarray}
Y1_{\mu}\,^{\nu} &=& 8\pi\theta_{\mu}\,^{\nu}\,,\\
Y2^{\lambda\mu\nu} &=& 4\pi\bigtriangleup^{\lambda\mu\nu},
\end{eqnarray}
with $\tilde{\mathcal{L}}_{Q}$ representing the Lagrangian density composed by the quadratic order corrections of the dynamical nonmetricity field in \eqref{TracelessLagrangian}.

Additionally, the Expression~\eqref{full_action} describes the full gravitational action of our model, depending on the dynamical parts of the torsion and nonmetricity tensors. Then, the corresponding vacuum field equations can be written as
\begin{eqnarray}
X1_{\mu}\,^{\nu} &=& 0\,,\\
X2^{\lambda\mu\nu} &=& 0\,,
\end{eqnarray}
where
\begin{eqnarray}
X1_{\mu}\,^{\nu} &=& 2\,G_{\mu}\,^{\nu}+16\pi\tilde{\mathcal{L}}\,\delta_{\mu}\,^{\nu}+3\left(1-2a_{2}\right)\left(g_{\mu\rho}\nabla_{\lambda}T^{\left[\lambda\nu\rho\right]}+N_{\lambda\rho\mu}T^{\left[\lambda\nu\rho\right]}\right)+2\left(f_{1}-2e_{1}\right)\tilde{R}^{\lambda}\,_{\lambda\sigma\mu}\tilde{R}^{\rho}\,_{\rho}\,^{\sigma\nu}
\nonumber\\
&&+d_{1}\left[\left(\tilde{R}^{\nu}\,_{\lambda\rho\mu}+\tilde{R}_{\lambda}\,^{\nu}\,_{\mu\rho}\right)\left(\tilde{R}^{\left[\lambda\rho\right]}+\hat{R}^{\left[\lambda\rho\right]}\right)+\left(\tilde{R}_{\lambda\mu}+\hat{R}_{\lambda\mu}\right)\left(\tilde{R}^{\left[\nu\lambda\right]}+\hat{R}^{\left[\nu\lambda\right]}\right)\right]
\nonumber\\
&&+\frac{d_{1}}{2}\tilde{R}_{\lambda\rho\sigma\mu}
\left(4\tilde{R}^{\left[\nu\sigma\right]\lambda\rho}-2\tilde{R}^{\left[\lambda\rho\right]\nu\sigma}-\tilde{R}^{\left[\lambda\nu\right]\rho\sigma}-\tilde{R}^{\left[\rho\sigma\right]\lambda\nu}-\tilde{R}^{\left[\rho\nu\right]\sigma\lambda}-\tilde{R}^{\left[\sigma\lambda\right]\rho\nu}\right)\,
\nonumber\\
&&+f_{1}\left[8\tilde{R}_{(\lambda\rho)\mu\sigma}\tilde{R}^{(\lambda\rho)\sigma\nu}-\left(\tilde{R}_{\lambda\mu}-\hat{R}_{\lambda\mu}\right)\left(\hat{R}^{(\lambda\nu)}-\tilde{R}^{(\lambda\nu)}\right)
-\left(\tilde{R}^{\nu}\,_{\lambda\rho\mu}-\tilde{R}_{\lambda}\,^{\nu}\,_{\mu\rho}\right)\left(\tilde{R}^{(\lambda\rho)}-\hat{R}^{(\lambda\rho)}\right)\right]\,,\label{Tensor_tetradeq}
\end{eqnarray}
\begin{eqnarray}
X2^{\lambda\mu\nu} &=& d_{1}\Bigl\{
\nabla_{\rho}\left[g^{\mu\nu}\left(\tilde{R}^{\left[\lambda\rho\right]}+\hat{R}^{\left[\lambda\rho\right]}\right)-g^{\lambda\nu}\left(\tilde{R}^{\left[\mu\rho\right]}+\hat{R}^{\left[\mu\rho\right]}\right)+g^{\lambda\rho}\left(\tilde{R}^{\left[\mu\nu\right]}+\hat{R}^{\left[\mu\nu\right]}\right)-g^{\mu\rho}\left(\tilde{R}^{\left[\lambda\nu\right]}+\hat{R}^{\left[\lambda\nu\right]}\right)\right]
\Bigr.\,
\nonumber\\
\Bigl.
&&+\,N^{\nu\lambda}\,_{\rho}\left(\tilde{R}^{[\mu\rho]}+\hat{R}^{[\mu\rho]}\right)+N^{\mu}\,_{\rho}\,^{\lambda}\left(\tilde{R}^{[\rho\nu]}+\hat{R}^{[\rho\nu]}\right)-N^{\rho\lambda}\,_{\rho}\left(\tilde{R}^{[\mu\nu]}+\hat{R}^{[\mu\nu]}\right)
\Bigr.
\nonumber\\
\Bigl.
&&+\,N^{\mu\nu}\,_{\rho}\left(\tilde{R}^{[\lambda\rho]}+\hat{R}^{[\lambda\rho]}\right)+N_{\rho}\,^{\lambda\mu}\left(\tilde{R}^{[\rho\nu]}+\hat{R}^{[\rho\nu]}\right)-N^{\mu\rho}\,_{\rho}\left(\tilde{R}^{[\lambda\nu]}+\hat{R}^{[\lambda\nu]}\right)
\Bigr\}\nonumber\\
&&+\,\frac{d_{1}}{2}\Bigl\{
\nabla_{\rho}\left(4\tilde{R}^{\left[\rho\nu\right]\lambda\mu}-2\tilde{R}^{\left[\lambda\mu\right]\rho\nu}-\tilde{R}^{\left[\mu\nu\right]\lambda\rho}+\tilde{R}^{\left[\lambda\nu\right]\mu\rho}-\tilde{R}^{\left[\lambda\rho\right]\mu\nu}+\tilde{R}^{\left[\mu\rho\right]\lambda\nu}\right)
\Bigr.\,
\nonumber\\
\Bigl.
&&+N^{\mu}\,_{\sigma\rho}\left[4\tilde{R}^{\left[\rho\nu\right]\lambda\sigma}-2\tilde{R}^{\left[\lambda\sigma\right]\rho\nu}-\tilde{R}^{\left[\sigma\nu\right]\lambda\rho}+\tilde{R}^{\left[\lambda\nu\right]\sigma\rho}-\tilde{R}^{\left[\lambda\rho\right]\sigma\nu}+\tilde{R}^{\left[\sigma\rho\right]\lambda\nu}-2g^{\lambda\nu}\left(\tilde{R}^{[\sigma\rho]}+\hat{R}^{[\sigma\rho]}\right)\right]
\Bigr.\,
\nonumber\\
\Bigl.
&&-N_{\sigma}\,^{\lambda}\,_{\rho}\left[4\tilde{R}^{\left[\rho\nu\right]\sigma\mu}-2\tilde{R}^{\left[\sigma\mu\right]\rho\nu}-\tilde{R}^{\left[\mu\nu\right]\sigma\rho}+\tilde{R}^{\left[\sigma\nu\right]\mu\rho}-\tilde{R}^{\left[\sigma\rho\right]\mu\nu}+\tilde{R}^{\left[\mu\rho\right]\sigma\nu}+2g^{\mu\nu}\left(\tilde{R}^{[\sigma\rho]}+\hat{R}^{[\sigma\rho]}\right)\right]
\Bigr\}\nonumber\\
&&+f_{1}\Bigl[
g^{\lambda\nu}\left(\nabla_{\rho}\tilde{R}^{(\mu\rho)}-\nabla_{\rho}\hat{R}^{(\mu\rho)}\right)+g^{\lambda\rho}\left(\nabla_{\rho}\hat{R}^{(\mu\nu)}-\nabla_{\rho}\tilde{R}^{(\mu\nu)}\right)+g^{\mu\nu}\left(\nabla_{\rho}\tilde{R}^{(\lambda\rho)}-\nabla_{\rho}\hat{R}^{(\lambda\rho)}\right)
\Bigr.\,
\nonumber\\
\Bigl.
&&+\,g^{\mu\rho}\left(\nabla_{\rho}\hat{R}^{(\lambda\nu)}-\nabla_{\rho}\tilde{R}^{(\lambda\nu)}\right)+N^{\mu\nu}\,_{\rho}\left(\tilde{R}^{(\lambda\rho)}-\hat{R}^{(\lambda\rho)}\right)+N^{\nu\lambda}\,_{\rho}\left(\hat{R}^{(\mu\rho)}-\tilde{R}^{(\mu\rho)}\right)+N^{\rho\lambda}\,_{\rho}\left(\tilde{R}^{(\mu\nu)}-\hat{R}^{(\mu\nu)}\right)
\Bigr.\,
\nonumber\\
\Bigl.
&&+\,\Bigl(g^{\lambda\nu}N^{\mu}\,_{\sigma\rho}-g^{\mu\nu}N_{\sigma}\,^{\lambda}\,_{\rho}\Bigr)\left(\tilde{R}^{(\sigma\rho)}-\hat{R}^{(\sigma\rho)}\right)+\Bigl(N_{\rho}\,^{\lambda\mu}-N^{\mu}\,_{\rho}\,^{\lambda}\Bigr)\left(\tilde{R}^{(\nu\rho)}-\hat{R}^{(\nu\rho)}\right)+N^{\mu\rho}\,_{\rho}\left(\hat{R}^{(\lambda\nu)}-\tilde{R}^{(\lambda\nu)}\right)\Bigr]
\nonumber\\
&&+\,8f_{1}\left(\nabla_{\rho}\tilde{R}^{(\mu\lambda)\rho\nu}+N^{\mu}\,_{\sigma\rho}\tilde{R}^{(\lambda\sigma)\rho\nu}-N_{\sigma}\,^{\lambda}\,_{\rho}\tilde{R}^{(\mu\sigma)\rho\nu}\right)+2\left(f_{1}-2e_{1}\right)g^{\lambda\mu}\nabla_{\rho}\tilde{R}^{\sigma}\,_{\sigma}\,^{\rho\nu}-3\left(1-2a_{2}\right)T^{[\lambda\mu\nu]}\,,\label{Tensor_connectioneq}
\end{eqnarray}
with $\tilde{\mathcal{L}}$ representing the Lagrangian density composed by all the quadratic order corrections in \eqref{full_action}.

\section{Irreducible parts of the complete solution}\label{sec:AppIrrPartsSol}

In this appendix, we show the torsion and nonmetricity irreducible parts for the complete Reissner-Nordstr\"{o}m-like solution presented in Sec.~\ref{sec:fullmodel}.

For the torsion sector, we separate the irreducible parts in dynamical and nondynamical quantities
\begin{eqnarray}
T_{\mu}&=&\mathring{T}_{\mu}\,,\\
S_{\mu}&=&\bar{S}_{\mu}\,,\\
t_{\lambda\mu\nu}&=&\mathring{t}_{\lambda\mu\nu}+\bar{t}_{\lambda\mu\nu}\,,
\end{eqnarray}
denoted by a bar and circle on top, respectively. The latter is in turn expressed as a sum of two contributions
\begin{eqnarray}
\mathring{T}_{\mu}&=&\mathring{T}_{1\mu}+\mathring{T}_{2\mu}\,,\\
\mathring{t}_{\lambda\mu\nu}&=&\mathring{t}_{1\lambda\mu\nu}+\mathring{t}_{2\lambda\mu\nu}\,,
\end{eqnarray}
which account for the part of the torsion tensor that provides the limit to GR in the absence of the rest of dynamical quantities of the solution, as well as for the part that preserves the components of the anholonomic connection associated with nonmetricity to be fully symmetric, namely
\begin{equation}
   \mathring{T}_{2\mu}=\frac{3}{2}\left(W_{\mu}-\,\frac{3}{4}\Lambda_{\mu}\right)\,, \quad \mathring{t}_{2\lambda\mu\nu}=\frac{1}{4}\varepsilon_{\sigma\rho\mu\nu}\Omega_\lambda{}^{\sigma\rho}\,.
\end{equation}

Then, the irreducible parts of the torsion tensor can be written as follows:
\begin{align}
    \mathring{T}_{1t}&=-\,\Psi\mathring{T}_{1r}=\frac{2\Psi+r\Psi'+4wr^{2}}{2r}\,, \quad \mathring{T}_{2t}=-\,\Psi\mathring{T}_{2r}=\frac{3\kappa_{\rm d}+\kappa_{\rm sh}-2r^{2}\bigl(\Psi t'_{\textrm{b},4}+t_{\textrm{b},4}\Psi'\bigr)-6rt_{\textrm{b},4}\Psi}{2r}\,,\\ \bar{S}_{t}&=-\,\Psi\bar{S}_{r}=-\,\frac{4\kappa_{\rm s}}{r}\,, \quad \mathring{t}_{1}^{r}{}_{t r}=\Psi\mathring{t}_{1}^{t}{}_{t r}=-\,2\mathring{t}_{1}^{\vartheta}{}_{t \vartheta}=-\,2\mathring{t}_{1}^{\varphi}{}_{t \varphi}=2\Psi\mathring{t}_{1}^{\vartheta}{}_{r \vartheta}=2\Psi\mathring{t}_{1}^{\varphi}{}_{r \varphi}=\frac{r\Psi'-\Psi+wr^{2}}{3r}\,,\\
    \mathring{t}_{2}^{r}{}_{t r}&=\Psi\mathring{t}_{2}^{t}{}_{t r}=-\,2\mathring{t}_{2}^{\vartheta}{}_{t \vartheta}=-\,2\mathring{t}_{2}^{\varphi}{}_{t \varphi}=2\Psi\mathring{t}_{2}^{\vartheta}{}_{r \vartheta}=2\Psi\mathring{t}_{2}^{\varphi}{}_{r \varphi}=\frac{\kappa_{\rm sh}-2r^{2}\bigl(\Psi t'_{\textrm{b},4}+t_{\textrm{b},4}\Psi'\bigr)}{3r}\,,\\
    \bar{t}^{\,\vartheta}{}_{t\varphi}&=-\,\Psi\bar{t}^{\,\vartheta}{}_{r\varphi}=-\,\bar{t}^{\,\varphi}{}_{t\vartheta}\sin^{2}\vartheta=\Psi\bar{t}^{\,\varphi}{}_{r\vartheta}\sin^{2}\vartheta=-\,\frac{1}{2r^{2}}\,\bar{t}^{\,r}{}_{\vartheta\varphi}=-\,\frac{\Psi}{2r^{2}}\,\bar{t}^{\,t}{}_{\vartheta\varphi}=-\,\frac{\kappa_{\rm s}\sin\vartheta}{3r}\,.
\end{align}

On the other hand, all the irreducible parts of nonmetricity act as dynamical quantities in the solution and read:
\begin{align}
    W_{t}&=-\,\Psi W_{r}=\frac{\kappa_{\rm d}}{r}\,, \quad \Lambda_{t}=-\,\Psi\Lambda_{r}=-\,\frac{4\bigl[\kappa_{\rm sh}-2r^{2}\bigl(\Psi t'_{\textrm{b},4}+t_{\textrm{b},4}\Psi'\bigr)-6rt_{\textrm{b},4}\Psi\bigr]}{9r}\,,\\
    \Omega_{\vartheta}{}^{r\varphi}&=\Psi\Omega_{\vartheta}{}^{t\varphi}=-\,\frac{1}{\sin^{2}\vartheta}\,\Omega_{\varphi}{}^{r\vartheta}=-\,\frac{\Psi}{\sin^{2}\vartheta}\,\Omega_{\varphi}{}^{t\vartheta}=-\,\frac{r^2}{2\,}\,\Omega_{t}{}^{\vartheta\varphi}=\frac{r^2\Psi}{2}\,\Omega_{r}{}^{\vartheta\varphi}=-\,\frac{\kappa_{\rm sh}-2r^{2}\bigl(\Psi t'_{\textrm{b},4}+t_{\textrm{b},4}\Psi'\bigr)}{3r\sin\vartheta}\,,\\
    q_{ttt}&=-\,\frac{\Psi\bigl[2\left(4\kappa_{\rm sh}+3c_{2}r\right)-4r^{2}\bigl(\Psi t'_{\textrm{b},4}+t_{\textrm{b},4}\Psi'\bigr)-3rq_{\textrm{b},7}\Psi\bigr]}{3r}\,,\\
    q_{ttr}&=\frac{14\kappa_{\rm sh}+9c_{2}r-10r^{2}\bigl(\Psi t'_{\textrm{b},4}+t_{\textrm{b},4}\Psi'\bigr)-9rq_{\textrm{b},7}\Psi-6rt_{\textrm{b},4}\Psi}{9r}\,,\\
    q_{trr}&=-\,\frac{4\kappa_{\rm sh}-8r^{2}\bigl(\Psi t'_{\textrm{b},4}+t_{\textrm{b},4}\Psi'\bigr)-9rq_{\textrm{b},7}\Psi-12rt_{\textrm{b},4}\Psi}{9r\Psi}\,,\\
    q_{rrr}&=-\,\frac{2\kappa_{\rm sh}+3c_{2}r+2r^{2}\bigl(\Psi t'_{\textrm{b},4}+t_{\textrm{b},4}\Psi'\bigr)+3rq_{\textrm{b},7}\Psi+6rt_{\textrm{b},4}\Psi}{3r\Psi^{2}}\,,\\
    q_{t\varphi\varphi}&=-\,\Psi q_{r\varphi\varphi}=q_{t\vartheta\vartheta}\sin^{2}\vartheta=-\,\Psi q_{r\vartheta\vartheta}\sin^{2}\vartheta=-\,\frac{r\sin^{2}\vartheta\bigl[10\kappa_{\rm sh}+9c_{2}r-2r^{2}\bigl(\Psi t'_{\textrm{b},4}+t_{\textrm{b},4}\Psi'\bigr)+6rt_{\textrm{b},4}\Psi\bigr]}{9}\,,
\end{align}
with $q_{\textrm{b},7}(r)=\frac{r^2}{\Psi^{2}(r)}\Big(c_4+\int\tilde{q}_{\textrm{b},7}(r) dr \Big)$, being $t_{\textrm{b},4}(r)$ and $\tilde{q}_{\textrm{b},7}(r)$ the expressions shown in Table~\ref{tab:final_functions} for the three different sets of the Lagrangian coefficients of the model.

\newpage

\bibliographystyle{utphys}
\bibliography{references}

\providecommand{\href}[2]{#2}\begingroup\raggedright\begin{thebibliography}{10}

\bibitem{Wald:1984rg}
R.~M. Wald,
  \href{http://dx.doi.org/10.7208/chicago/9780226870373.001.0001}{{\em {General
  Relativity}}}.
\newblock Chicago Univ. Pr., Chicago, USA, 1984.

\bibitem{Hehl:1976kj}
F.~Hehl, P.~von~der Heyde, G.~Kerlick, and J.~Nester, ``{General Relativity
  with Spin and Torsion: Foundations and Prospects},''
  \href{http://dx.doi.org/10.1103/RevModPhys.48.393}{{\em Rev. Mod. Phys.} {\bf
  48} (1976)  393--416}.

\bibitem{Obukhov:1987tz}
Y.~Obukhov, V.~Ponomarev, and V.~Zhytnikov, ``{Quadratic Poincar\'e Gauge
  Theory of Gravity: A Comparison With the General Relativity Theory},''
  \href{http://dx.doi.org/10.1007/BF00763457}{{\em Gen. Rel. Grav.} {\bf 21}
  (1989)  1107--1142}.

\bibitem{Blagojevic:2013xpa}
M.~Blagojevi\'c and F.~W. Hehl, eds., {\em {Gauge Theories of Gravitation}: {A
  Reader with Commentaries}}.
\newblock World Scientific, Singapore, 2013.

\bibitem{Obukhov:2022khx}
Y.~N. Obukhov, ``{Poincar\'e gauge gravity primer},''
  \href{http://arxiv.org/abs/2206.05205}{{\tt arXiv:2206.05205 [gr-qc]}}.

\bibitem{Hehl:1994ue}
F.~W. Hehl, J.~D. McCrea, E.~W. Mielke, and Y.~Ne'eman, ``{Metric-Affine gauge
  theory of gravity: Field equations, Noether identities, world spinors, and
  breaking of dilation invariance},''
  \href{http://dx.doi.org/10.1016/0370-1573(94)00111-F}{{\em Phys. Rept.} {\bf
  258} (1995)  1--171},
\href{http://arxiv.org/abs/gr-qc/9402012}{{\tt arXiv:gr-qc/9402012 [gr-qc]}}.
%%CITATION = GR-QC/9402012;%%.

\bibitem{ponomarev2017gauge}
{V. N. Ponomarev, A. O. Barvinsky, and Yu. N. Obukhov}, {\em Gauge Approach and
  Quantization Methods in Gravity Theory}.
\newblock Nauka, Moscow, 2017.

\bibitem{Cabral:2020fax}
F.~Cabral, F.~S. Lobo, and D.~Rubiera-Garcia, ``{Fundamental Symmetries and
  Spacetime Geometries in Gauge Theories of Gravity: Prospects for Unified
  Field Theories},'' \href{http://dx.doi.org/10.3390/universe6120238}{{\em
  Universe} {\bf 6} (2020) no.~12, 238},
  \href{http://arxiv.org/abs/2012.06356}{{\tt arXiv:2012.06356 [gr-qc]}}.

\bibitem{Bakler:1985qj}
P.~Baekler, F.~W. Hehl, and E.~W. Mielke, ``{Nonmetricity and torsion: facts
  and fancies in gauge approaches to gravity},'' in {\em {4th Marcel Grossmann
  Meeting on the Recent Developments of General Relativity}}.
\newblock 6, 1985.

\bibitem{Gronwald:1995em}
F.~Gronwald and F.~W. Hehl, ``{On the gauge aspects of gravity},'' in {\em
  {International School of Cosmology and Gravitation: 14th Course: Quantum
  Gravity}}, pp.~148--198.
\newblock 5, 1995.
\newblock \href{http://arxiv.org/abs/gr-qc/9602013}{{\tt arXiv:gr-qc/9602013}}.

\bibitem{adamowicz1980plane}
W.~Adamowicz, ``Plane waves in gauge theories of gravitation,'' {\em General
  Relativity and Gravitation} {\bf 12} (1980) no.~9, 677--691.

\bibitem{Baekler:1981lkh}
P.~Baekler, ``{A spherically symmetric vacuum solution of the quadratic
  Poincar\'e gauge field theory of gravitation with newtonian and confinement
  potentials},'' \href{http://dx.doi.org/10.1016/0370-2693(81)90111-8}{{\em
  Phys. Lett. B} {\bf 99} (1981)  329--332}.

\bibitem{Gonner:1984rw}
H.~Gonner and F.~Mueller-Hoissen, ``{Spatially homogeneous and isotropic spaces
  in theories of gravitation with torsion},''
  \href{http://dx.doi.org/10.1088/0264-9381/1/6/010}{{\em Class. Quant. Grav.}
  {\bf 1} (1984)  651}.

\bibitem{Bakler:1988nq}
P.~Baekler, M.~Gurses, F.~W. Hehl, and J.~D. McCrea, ``{The exterior
  gravitational field of a charged spinning source in the Poincar\'{e} Gauge
  theory: A Kerr-Newman metric with dynamic torsion},''
  \href{http://dx.doi.org/10.1016/0375-9601(88)90366-0}{{\em Phys. Lett. A}
  {\bf 128} (1988)  245--250}.

\bibitem{Tresguerres:1995js}
R.~Tresguerres, ``{Exact vacuum solutions of four-dimensional metric-affine
  gauge theories of gravitation},''
  \href{http://dx.doi.org/10.1007/BF01571892}{{\em Z. Phys. C} {\bf 65} (1995)
  347--354}.

\bibitem{tresguerres1995exact}
R.~Tresguerres, ``Exact static vacuum solution of four-dimensional
  metric-affine gravity with nontrivial torsion,'' {\em Physics Letters A} {\bf
  200} (1995) no.~6, 405--410.

\bibitem{Hehl:1999sb}
F.~W. Hehl and A.~Macias, ``{Metric-Affine gauge theory of gravity. 2. Exact
  solutions},'' \href{http://dx.doi.org/10.1142/S0218271899000316}{{\em Int. J.
  Mod. Phys. D} {\bf 8} (1999)  399--416},
  \href{http://arxiv.org/abs/gr-qc/9902076}{{\tt arXiv:gr-qc/9902076}}.

\bibitem{Garcia:2000yi}
A.~Garcia, A.~Macias, D.~Puetzfeld, and J.~Socorro, ``{Plane fronted waves in
  metric-affine gravity},''
  \href{http://dx.doi.org/10.1103/PhysRevD.62.044021}{{\em Phys. Rev. D} {\bf
  62} (2000)  044021}, \href{http://arxiv.org/abs/gr-qc/0005038}{{\tt
  arXiv:gr-qc/0005038}}.

\bibitem{Puetzfeld:2001hk}
D.~Puetzfeld and R.~Tresguerres, ``{A cosmological model in Weyl-Cartan
  spacetime},'' \href{http://dx.doi.org/10.1088/0264-9381/18/4/308}{{\em Class.
  Quant. Grav.} {\bf 18} (2001)  677--694},
  \href{http://arxiv.org/abs/gr-qc/0101050}{{\tt arXiv:gr-qc/0101050}}.

\bibitem{Chen:2009at}
H.~Chen, F.-H. Ho, J.~M. Nester, C.-H. Wang, and H.-J. Yo, ``{Cosmological
  dynamics with propagating Lorentz connection modes of spin zero},''
  \href{http://dx.doi.org/10.1088/1475-7516/2009/10/027}{{\em JCAP} {\bf 10}
  (2009)  027}, \href{http://arxiv.org/abs/0908.3323}{{\tt arXiv:0908.3323
  [gr-qc]}}.

\bibitem{Lu:2016bcx}
J.~Lu and G.~Chee, ``{Cosmology in Poincar\'e gauge gravity with a pseudoscalar
  torsion},'' \href{http://dx.doi.org/10.1007/JHEP05(2016)024}{{\em JHEP} {\bf
  05} (2016)  024}, \href{http://arxiv.org/abs/1601.03943}{{\tt
  arXiv:1601.03943 [gr-qc]}}.

\bibitem{Boos:2016cey}
J.~Boos and F.~W. Hehl, ``{Gravity-induced four-fermion contact interaction
  implies gravitational intermediate W and Z type gauge bosons},''
  \href{http://dx.doi.org/10.1007/s10773-016-3216-3}{{\em Int. J. Theor. Phys.}
  {\bf 56} (2017) no.~3, 751--756}, \href{http://arxiv.org/abs/1606.09273}{{\tt
  arXiv:1606.09273 [gr-qc]}}.

\bibitem{Cembranos:2016gdt}
J.~A.~R. Cembranos and J.~Gigante~Valcarcel, ``{New torsion black hole
  solutions in Poincar\'e gauge theory},''
  \href{http://dx.doi.org/10.1088/1475-7516/2017/01/014}{{\em JCAP} {\bf 01}
  (2017)  014}, \href{http://arxiv.org/abs/1608.00062}{{\tt arXiv:1608.00062
  [gr-qc]}}.

\bibitem{Cembranos:2017pcs}
J.~A.~R. Cembranos and J.~Gigante~Valcarcel, ``{Extended Reissner--Nordström
  solutions sourced by dynamical torsion},''
  \href{http://dx.doi.org/10.1016/j.physletb.2018.01.081}{{\em Phys. Lett. B}
  {\bf 779} (2018)  143--150}, \href{http://arxiv.org/abs/1708.00374}{{\tt
  arXiv:1708.00374 [gr-qc]}}.

\bibitem{Blagojevic:2017wzf}
M.~Blagojevi\'c and B.~Cvetkovi\'c, ``{Generalized pp waves in Poincar\'e gauge
  theory},'' \href{http://dx.doi.org/10.1103/PhysRevD.95.104018}{{\em Phys.
  Rev. D} {\bf 95} (2017) no.~10, 104018},
  \href{http://arxiv.org/abs/1702.04367}{{\tt arXiv:1702.04367 [gr-qc]}}.

\bibitem{Blagojevic:2017ssv}
M.~Blagojevi\'c, B.~Cvetkovi\'c, and Y.~N. Obukhov, ``{Generalized plane waves
  in Poincar\'e gauge theory of gravity},''
  \href{http://dx.doi.org/10.1103/PhysRevD.96.064031}{{\em Phys. Rev. D} {\bf
  96} (2017) no.~6, 064031}, \href{http://arxiv.org/abs/1708.08766}{{\tt
  arXiv:1708.08766 [gr-qc]}}.

\bibitem{Obukhov:2019fti}
Y.~N. Obukhov, ``{Exact Solutions in Poincar\'e Gauge Gravity Theory},''
  \href{http://dx.doi.org/10.3390/universe5050127}{{\em Universe} {\bf 5}
  (2019) no.~5, 127}, \href{http://arxiv.org/abs/1905.11906}{{\tt
  arXiv:1905.11906 [gr-qc]}}.

\bibitem{Guerrero:2020azx}
M.~Guerrero, G.~Mora-P\'erez, G.~J. Olmo, E.~Orazi, and D.~Rubiera-Garcia,
  ``{Rotating black holes in Eddington-inspired Born-Infeld gravity: an exact
  solution},'' \href{http://dx.doi.org/10.1088/1475-7516/2020/07/058}{{\em
  JCAP} {\bf 07} (2020)  058}, \href{http://arxiv.org/abs/2006.00761}{{\tt
  arXiv:2006.00761 [gr-qc]}}.

\bibitem{Bahamonde:2020fnq}
S.~Bahamonde and J.~Gigante~Valcarcel, ``{New models with independent dynamical
  torsion and nonmetricity fields},''
  \href{http://dx.doi.org/10.1088/1475-7516/2020/09/057}{{\em JCAP} {\bf 09}
  (2020)  057}, \href{http://arxiv.org/abs/2006.06749}{{\tt arXiv:2006.06749
  [gr-qc]}}.

\bibitem{Obukhov:2020hlp}
Y.~N. Obukhov, ``{Generalized Birkhoff theorem in the Poincar\'e gauge gravity
  theory},'' \href{http://dx.doi.org/10.1103/PhysRevD.102.104059}{{\em Phys.
  Rev. D} {\bf 102} (2020) no.~10, 104059},
  \href{http://arxiv.org/abs/2009.00284}{{\tt arXiv:2009.00284 [gr-qc]}}.

\bibitem{Iosifidis:2020gth}
D.~Iosifidis, ``{Cosmological hyperfluids, torsion and non-metricity},''
  \href{http://dx.doi.org/10.1140/epjc/s10052-020-08634-z}{{\em Eur. Phys. J.
  C} {\bf 80} (2020) no.~11, 1042}, \href{http://arxiv.org/abs/2003.07384}{{\tt
  arXiv:2003.07384 [gr-qc]}}.

\bibitem{Aoki:2020zqm}
K.~Aoki and S.~Mukohyama, ``{Consistent inflationary cosmology from quadratic
  gravity with dynamical torsion},''
  \href{http://dx.doi.org/10.1088/1475-7516/2020/06/004}{{\em JCAP} {\bf 06}
  (2020)  004}, \href{http://arxiv.org/abs/2003.00664}{{\tt arXiv:2003.00664
  [hep-th]}}.

\bibitem{Bahamonde:2021qjk}
S.~Bahamonde and J.~Gigante~Valcarcel, ``{Rotating Kerr-Newman space-times in
  metric-affine gravity},''
  \href{http://dx.doi.org/10.1088/1475-7516/2022/01/011}{{\em JCAP} {\bf 01}
  (2022) no.~01, 011}, \href{http://arxiv.org/abs/2108.12414}{{\tt
  arXiv:2108.12414 [gr-qc]}}.

\bibitem{delaCruzDombriz:2021nrg}
{A. de la Cruz-Dombriz, F. J. Maldonado Torralba, and D. F. Mota}, ``{Dark
  matter candidate from torsion},''
  \href{http://dx.doi.org/10.1016/j.physletb.2022.137488}{{\em Phys. Lett. B}
  {\bf 834} (2022)  137488}, \href{http://arxiv.org/abs/2112.03957}{{\tt
  arXiv:2112.03957 [gr-qc]}}.

\bibitem{Jimenez-Cano:2022arz}
A.~Jim\'enez-Cano, ``{Review of gravitational wave solutions in quadratic
  metric-affine gravity},'' in {\em {2022 Snowmass Summer Study}}.
\newblock 3, 2022.
\newblock \href{http://arxiv.org/abs/2203.03936}{{\tt arXiv:2203.03936
  [gr-qc]}}.

\bibitem{Bahamonde:2022meb}
S.~Bahamonde, J.~Gigante~Valcarcel, and L.~Järv, ``{Pleba\'nski-Demia\'nski
  solutions with dynamical torsion and nonmetricity fields},''
  \href{http://dx.doi.org/10.1088/1475-7516/2022/04/011}{{\em JCAP} {\bf 04}
  (2022) no.~04, 011}, \href{http://arxiv.org/abs/2201.10532}{{\tt
  arXiv:2201.10532 [gr-qc]}}.

\bibitem{Bombacigno:2022lcx}
F.~Bombacigno, ``{Quasinormal modes of Schwarzschild black holes in metric
  affine Chern-Simons theory},'' in {\em {56th Rencontres de Moriond on
  Gravitation}}.
\newblock 3, 2022.
\newblock \href{http://arxiv.org/abs/2203.09209}{{\tt arXiv:2203.09209
  [gr-qc]}}.

\bibitem{Boudet:2022nub}
S.~Boudet, F.~Bombacigno, F.~Moretti, and G.~J. Olmo, ``{Torsional
  birefringence in metric-affine Chern-Simons gravity: gravitational waves in
  late-time cosmology},''
  \href{http://dx.doi.org/10.1088/1475-7516/2023/01/026}{{\em JCAP} {\bf 01}
  (2023)  026}, \href{http://arxiv.org/abs/2209.14394}{{\tt arXiv:2209.14394
  [gr-qc]}}.

\bibitem{Yang:2022icz}
J.-Z. Yang, S.~Shahidi, and T.~Harko, ``{Black hole solutions in the quadratic
  Weyl conformal geometric theory of gravity},''
  \href{http://dx.doi.org/10.1140/epjc/s10052-022-11131-0}{{\em Eur. Phys. J.
  C} {\bf 82} (2022) no.~12, 1171}, \href{http://arxiv.org/abs/2212.05542}{{\tt
  arXiv:2212.05542 [gr-qc]}}.

\bibitem{Iosifidis:2022xvp}
D.~Iosifidis, R.~Myrzakulov, and L.~Ravera, ``{Cosmology of Metric-Affine $R +
  \beta R^2$ Gravity with Pure Shear Hypermomentum},''
  \href{http://arxiv.org/abs/2301.00669}{{\tt arXiv:2301.00669 [gr-qc]}}.

\bibitem{Heinicke:2005bp}
C.~Heinicke, P.~Baekler, and F.~W. Hehl, ``{Einstein-aether theory, violation
  of Lorentz invariance, and metric-affine gravity},''
  \href{http://dx.doi.org/10.1103/PhysRevD.72.025012}{{\em Phys. Rev. D} {\bf
  72} (2005)  025012}, \href{http://arxiv.org/abs/gr-qc/0504005}{{\tt
  arXiv:gr-qc/0504005}}.

\bibitem{Baekler:2006de}
P.~Baekler and F.~W. Hehl, ``{Rotating black holes in metric-affine gravity},''
  \href{http://dx.doi.org/10.1142/S0218271806008589}{{\em Int. J. Mod. Phys. D}
  {\bf 15} (2006)  635--668}, \href{http://arxiv.org/abs/gr-qc/0601063}{{\tt
  arXiv:gr-qc/0601063}}.

\bibitem{Dadhich:2010xa}
N.~Dadhich and J.~M. Pons, ``{On the equivalence of the Einstein-Hilbert and
  the Einstein-Palatini formulations of General Relativity for an arbitrary
  connection},'' \href{http://dx.doi.org/10.1007/s10714-012-1393-9}{{\em Gen.
  Rel. Grav.} {\bf 44} (2012)  2337--2352},
  \href{http://arxiv.org/abs/1010.0869}{{\tt arXiv:1010.0869 [gr-qc]}}.

\bibitem{BeltranJimenez:2017doy}
J.~Beltran~Jimenez, L.~Heisenberg, G.~J. Olmo, and D.~Rubiera-Garcia,
  ``{Born–Infeld inspired modifications of gravity},''
  \href{http://dx.doi.org/10.1016/j.physrep.2017.11.001}{{\em Phys. Rept.} {\bf
  727} (2018)  1--129},
\href{http://arxiv.org/abs/1704.03351}{{\tt arXiv:1704.03351 [gr-qc]}}.
%%CITATION = ARXIV:1704.03351;%%.

\bibitem{Afonso:2017bxr}
V.~I. Afonso, C.~Bejarano, J.~Beltran~Jimenez, G.~J. Olmo, and E.~Orazi, ``{The
  trivial role of torsion in projective invariant theories of gravity with
  non-minimally coupled matter fields},''
  \href{http://dx.doi.org/10.1088/1361-6382/aa9151}{{\em Class. Quant. Grav.}
  {\bf 34} (2017) no.~23, 235003}, \href{http://arxiv.org/abs/1705.03806}{{\tt
  arXiv:1705.03806 [gr-qc]}}.

\bibitem{Sadovski:2022kwf}
G.~Sadovski, ``{About the (in)equivalence between holonomic versus
  non-holonomic theories of gravity},''
  \href{http://arxiv.org/abs/2207.05721}{{\tt arXiv:2207.05721 [gr-qc]}}.

\bibitem{McCrea:1992wa}
J.~D. McCrea, ``{Irreducible decompositions of non-metricity, torsion,
  curvature and Bianchi identities in metric-affine spacetimes},''
  \href{http://dx.doi.org/10.1088/0264-9381/9/2/018}{{\em Class. Quant. Grav.}
  {\bf 9} (1992)  553--568}.

\bibitem{Hehl:2007bn}
F.~W. Hehl and Y.~N. Obukhov, ``{Elie Cartan's torsion in geometry and in field
  theory, an essay},'' {\em Annales Fond. Broglie} {\bf 32} (2007)  157--194,
  \href{http://arxiv.org/abs/0711.1535}{{\tt arXiv:0711.1535 [gr-qc]}}.

\bibitem{gambini1980einstein}
R.~Gambini and L.~Herrera, ``{Einstein-Cartan theory in the spin coefficient
  formalism},'' {\em Journal of Mathematical Physics} {\bf 21} (1980) no.~6,
  1449--1454.

\bibitem{Cembranos:2018ipn}
{J. A. R. Cembranos, J. Gigante Valcarcel, and F. J. Maldonado Torralba},
  ``{Fermion dynamics in torsion theories},''
  \href{http://dx.doi.org/10.1088/1475-7516/2019/04/039}{{\em JCAP} {\bf 04}
  (2019)  039}, \href{http://arxiv.org/abs/1805.09577}{{\tt arXiv:1805.09577
  [gr-qc]}}.

\bibitem{tsamparlis1979cosmological}
M.~Tsamparlis, ``Cosmological principle and torsion,'' {\em Physics Letters A}
  {\bf 75} (1979) no.~1-2, 27--28.

\bibitem{Puetzfeld:2004yg}
D.~Puetzfeld, ``{Status of non-Riemannian cosmology},''
  \href{http://dx.doi.org/10.1016/j.newar.2005.01.022}{{\em New Astron. Rev.}
  {\bf 49} (2005)  59--64}, \href{http://arxiv.org/abs/gr-qc/0404119}{{\tt
  arXiv:gr-qc/0404119}}.

\bibitem{Neeman:1978jlt}
Y.~Ne'eman and D.~Sijacki, ``{Unified Affine Gauge Theory of Gravity and Strong
  Interactions with Finite and Infinite $\overline{GL}(4,R)$ Spinor Fields},''
  \href{http://dx.doi.org/10.1016/0003-4916(79)90392-0}{{\em Annals Phys.} {\bf
  120} (1979)  292}. [Erratum: Annals Phys. 125, 227 (1980)].

\bibitem{Peterson:2019uzn}
C.~Peterson and Y.~Bonder, ``{Conserved quantities in the presence of torsion:
  A generalization of Killing theorem},''
  \href{http://dx.doi.org/10.1142/S0217732320500522}{{\em Mod. Phys. Lett. A}
  {\bf 35} (2019) no.~09, 2050052}, \href{http://arxiv.org/abs/1904.12913}{{\tt
  arXiv:1904.12913 [gr-qc]}}.

\bibitem{Hohmann:2019fvf}
M.~Hohmann, ``{Metric-Affine Geometries with Spherical Symmetry},''
  \href{http://dx.doi.org/10.3390/sym12030453}{{\em Symmetry} {\bf 12} (2020)
  no.~3, 453}, \href{http://arxiv.org/abs/1912.12906}{{\tt arXiv:1912.12906
  [math-ph]}}.

\bibitem{Lenzen:1986hb}
H.~Lenzen, ``{On Spherically Symmetric Fields with Dynamic Torsion in Gauge
  Theories of Gravitation},'' \href{http://dx.doi.org/10.1007/BF00773620}{{\em
  Gen. Rel. Grav.} {\bf 17} (1985)  1137--1151}.

\bibitem{chen1994poincare}
C.-M. Chen, D.-C. Chern, J.~M. Nester, P.-K. Yang, and V.~V. Zhytnikov,
  ``{Poincar{\'e} gauge theory Schwarzschild-de Sitter solutions with long
  range spherically symmetric torsion},'' {\em Chinese Journal of Physics} {\bf
  32} (1994) no.~1, 29--40.

\bibitem{ho1997some}
J.~Ho, D.-C. Chern, and J.~M. Nester, ``{Some Spherically Symmetric Exact
  Solutions of the Metric-Affine Gravity Theory},'' {\em Chin. J. Phys} {\bf
  35} (1997)  6--1.

\bibitem{hehl1991spacetime}
F.~W. Hehl and Y.~Ne'eman, ``Spacetime as a continuum with microstructure and
  metric-affine gravity,'' in {\em Modern Problems Of Theoretical Physics:
  Jubilee Volume of D Ivanenko's 85th Birthday}, pp.~31--52.
\newblock World Scientific, 1991.

\bibitem{Dothan:1965aa}
Y.~Dothan, M.~Gell-Mann, and Y.~Ne'eman, ``{Series of hadron energy levels as
  representations of non-compact groups},''
  \href{http://dx.doi.org/10.1016/0031-9163(65)90279-9}{{\em Phys. Lett.} {\bf
  17} (1965)  148--151}.

\bibitem{Neeman:1987nsi}
Y.~Ne'eman and D.~Sijacki, ``{Hadrons in an $\overline{SL}(4,R)$
  Classification: Phenomenology},''
  \href{http://dx.doi.org/10.1103/PhysRevD.37.3267}{{\em Phys. Rev. D} {\bf 37}
  (1988)  3267}.

\bibitem{Yagi:2016jml}
K.~Yagi and L.~C. Stein, ``{Black Hole Based Tests of General Relativity},''
  \href{http://dx.doi.org/10.1088/0264-9381/33/5/054001}{{\em Class. Quant.
  Grav.} {\bf 33} (2016) no.~5, 054001},
  \href{http://arxiv.org/abs/1602.02413}{{\tt arXiv:1602.02413 [gr-qc]}}.

\bibitem{Bambi:2015kza}
C.~Bambi, ``{Testing black hole candidates with electromagnetic radiation},''
  \href{http://dx.doi.org/10.1103/RevModPhys.89.025001}{{\em Rev. Mod. Phys.}
  {\bf 89} (2017) no.~2, 025001}, \href{http://arxiv.org/abs/1509.03884}{{\tt
  arXiv:1509.03884 [gr-qc]}}.

\bibitem{Berti:2018cxi}
E.~Berti, K.~Yagi, and N.~Yunes, ``{Extreme Gravity Tests with Gravitational
  Waves from Compact Binary Coalescences: (I) Inspiral-Merger},''
  \href{http://dx.doi.org/10.1007/s10714-018-2362-8}{{\em Gen. Rel. Grav.} {\bf
  50} (2018) no.~4, 46}, \href{http://arxiv.org/abs/1801.03208}{{\tt
  arXiv:1801.03208 [gr-qc]}}.

\bibitem{Barack:2018yly}
L.~Barack {\em et al.}, ``{Black holes, gravitational waves and fundamental
  physics: a roadmap},'' \href{http://dx.doi.org/10.1088/1361-6382/ab0587}{{\em
  Class. Quant. Grav.} {\bf 36} (2019) no.~14, 143001},
  \href{http://arxiv.org/abs/1806.05195}{{\tt arXiv:1806.05195 [gr-qc]}}.

\bibitem{Bahamonde:2021akc}
S.~Bahamonde and J.~Gigante~Valcarcel, ``{Observational constraints in
  metric-affine gravity},''
  \href{http://dx.doi.org/10.1140/epjc/s10052-021-09275-6}{{\em Eur. Phys. J.
  C} {\bf 81} (2021) no.~6, 495}, \href{http://arxiv.org/abs/2103.12036}{{\tt
  arXiv:2103.12036 [gr-qc]}}.

\bibitem{EventHorizonTelescope:2021dqv}
{P. Kocherlakota {\it et al.} (The Event Horizon Telescope Collaboration)},
  ``{Constraints on black-hole charges with the 2017 EHT observations of
  M87*},'' \href{http://dx.doi.org/10.1103/PhysRevD.103.104047}{{\em Phys. Rev.
  D} {\bf 103} (2021) no.~10, 104047},
  \href{http://arxiv.org/abs/2105.09343}{{\tt arXiv:2105.09343 [gr-qc]}}.

\bibitem{Kramer:2021jcw}
M.~Kramer {\em et al.}, ``{Strong-Field Gravity Tests with the Double
  Pulsar},'' \href{http://dx.doi.org/10.1103/PhysRevX.11.041050}{{\em Phys.
  Rev. X} {\bf 11} (2021) no.~4, 041050},
  \href{http://arxiv.org/abs/2112.06795}{{\tt arXiv:2112.06795 [astro-ph.HE]}}.

\bibitem{Daas:2022iid}
J.~Daas, K.~Kuijpers, F.~Saueressig, M.~F. Wondrak, and H.~Falcke, ``{Probing
  Quadratic Gravity with the Event Horizon Telescope},''
  \href{http://arxiv.org/abs/2204.08480}{{\tt arXiv:2204.08480 [gr-qc]}}.

\bibitem{EventHorizonTelescope:2022xqj}
{K. Akiyama {\it et al.} (The Event Horizon Telescope Collaboration)}, ``{First
  Sagittarius A* Event Horizon Telescope Results. VI. Testing the Black Hole
  Metric},'' \href{http://dx.doi.org/10.3847/2041-8213/ac6756}{{\em Astrophys.
  J. Lett.} {\bf 930} (2022) no.~2, L17}.

\bibitem{Pantig:2022ely}
R.~C. Pantig and A.~\"Ovg\"un, ``{Testing dynamical torsion effects on the
  charged black hole\textquoteright{}s shadow, deflection angle and greybody
  with M87* and Sgr. A* from EHT},''
  \href{http://dx.doi.org/10.1016/j.aop.2022.169197}{{\em Annals Phys.} {\bf
  448} (2023)  169197}, \href{http://arxiv.org/abs/2206.02161}{{\tt
  arXiv:2206.02161 [gr-qc]}}.

\bibitem{Jimenez:2019qjc}
J.~Beltrán~Jiménez and F.~J. Maldonado~Torralba, ``{Revisiting the stability
  of quadratic Poincaré gauge gravity},''
  \href{http://dx.doi.org/10.1140/epjc/s10052-020-8163-8}{{\em Eur. Phys. J. C}
  {\bf 80} (2020) no.~7, 611}, \href{http://arxiv.org/abs/1910.07506}{{\tt
  arXiv:1910.07506 [gr-qc]}}.

\bibitem{Percacci:2020ddy}
R.~Percacci and E.~Sezgin, ``{New class of ghost- and tachyon-free metric
  affine gravities},''
  \href{http://dx.doi.org/10.1103/PhysRevD.101.084040}{{\em Phys. Rev. D} {\bf
  101} (2020) no.~8, 084040}, \href{http://arxiv.org/abs/1912.01023}{{\tt
  arXiv:1912.01023 [hep-th]}}.

\bibitem{BeltranJimenez:2020sqf}
J.~Beltr\'an~Jim\'enez and A.~Delhom, ``{Instabilities in metric-affine
  theories of gravity with higher order curvature terms},''
  \href{http://dx.doi.org/10.1140/epjc/s10052-020-8143-z}{{\em Eur. Phys. J. C}
  {\bf 80} (2020) no.~6, 585}, \href{http://arxiv.org/abs/2004.11357}{{\tt
  arXiv:2004.11357 [gr-qc]}}.

\bibitem{Baldazzi:2021kaf}
A.~Baldazzi, O.~Melichev, and R.~Percacci, ``{Metric-Affine Gravity as an
  effective field theory},''
  \href{http://dx.doi.org/10.1016/j.aop.2022.168757}{{\em Annals Phys.} {\bf
  438} (2022)  168757}, \href{http://arxiv.org/abs/2112.10193}{{\tt
  arXiv:2112.10193 [gr-qc]}}.

\bibitem{Jimenez-Cano:2022sds}
A.~Jim\'enez-Cano and F.~J. Maldonado~Torralba, ``{Vector stability in
  quadratic metric-affine theories},''
  \href{http://dx.doi.org/10.1088/1475-7516/2022/09/044}{{\em JCAP} {\bf 09}
  (2022)  044}, \href{http://arxiv.org/abs/2205.05674}{{\tt arXiv:2205.05674
  [gr-qc]}}.

\bibitem{Blagojevic:2019gsd}
M.~Blagojevi\'c and B.~Cvetkovi\'c, ``{Entropy in Poincar\'e gauge theory:
  Hamiltonian approach},''
  \href{http://dx.doi.org/10.1103/PhysRevD.99.104058}{{\em Phys. Rev. D} {\bf
  99} (2019) no.~10, 104058}, \href{http://arxiv.org/abs/1903.02263}{{\tt
  arXiv:1903.02263 [gr-qc]}}.

\bibitem{Blagojevic:2021pqp}
M.~Blagojevi\'c and B.~Cvetkovi\'c, ``{Entropy of Reissner-Nordstr\"om-like
  black holes},'' \href{http://dx.doi.org/10.1016/j.physletb.2021.136815}{{\em
  Phys. Lett. B} {\bf 824} (2022)  136815},
  \href{http://arxiv.org/abs/2112.02099}{{\tt arXiv:2112.02099 [gr-qc]}}.

\bibitem{Blagojevic:2019bqg}
M.~Blagojevi\'c and B.~Cvetkovi\'c, ``{Hamiltonian approach to black hole
  entropy: Kerr-like spacetimes},''
  \href{http://dx.doi.org/10.1103/PhysRevD.100.044029}{{\em Phys. Rev. D} {\bf
  100} (2019) no.~4, 044029}, \href{http://arxiv.org/abs/1905.04928}{{\tt
  arXiv:1905.04928 [gr-qc]}}.

\bibitem{Blagojevic:2020edq}
M.~Blagojevi\'c and B.~Cvetkovi\'c, ``{Entropy in general relativity: Kerr-AdS
  black hole},'' \href{http://dx.doi.org/10.1103/PhysRevD.101.084023}{{\em
  Phys. Rev. D} {\bf 101} (2020) no.~8, 084023},
  \href{http://arxiv.org/abs/2002.05029}{{\tt arXiv:2002.05029 [gr-qc]}}.

\bibitem{Blagojevic:2020ymf}
M.~Blagojevi\'c and B.~Cvetkovi\'c, ``{Entropy in Poincar\'e gauge theory:
  Kerr-AdS solution},''
  \href{http://dx.doi.org/10.1103/PhysRevD.102.064034}{{\em Phys. Rev. D} {\bf
  102} (2020) no.~6, 064034}, \href{http://arxiv.org/abs/2007.10721}{{\tt
  arXiv:2007.10721 [gr-qc]}}.

\bibitem{Blagojevic:2021mli}
M.~Blagojevi\'c and B.~Cvetkovi\'c, ``{Thermodynamics of Riemannian Kerr-AdS
  black holes in Poincar\'e gauge theory},''
  \href{http://dx.doi.org/10.1016/j.physletb.2021.136242}{{\em Phys. Lett. B}
  {\bf 816} (2021)  136242}, \href{http://arxiv.org/abs/2103.00330}{{\tt
  arXiv:2103.00330 [gr-qc]}}.

\bibitem{Blagojevic:2022etm}
M.~Blagojevi\'c and B.~Cvetkovi\'c, ``{Entropy of Kerr-Newman-AdS black holes
  with torsion},'' \href{http://dx.doi.org/10.1103/PhysRevD.105.104014}{{\em
  Phys. Rev. D} {\bf 105} (2022) no.~10, 104014},
  \href{http://arxiv.org/abs/2203.14696}{{\tt arXiv:2203.14696 [gr-qc]}}.

\bibitem{Cvetkovic:2022qpt}
B.~Cvetkovi\'c and D.~Rakonjac, ``{Extremal Kerr black hole entropy in
  Poincar\'e gauge theory},'' \href{http://arxiv.org/abs/2208.04383}{{\tt
  arXiv:2208.04383 [gr-qc]}}.

\bibitem{Campos:1998xq}
A.~Campos and B.~L. Hu, ``{Fluctuations in a thermal field and dissipation of a
  black hole space-time: Far field limit},''
  \href{http://dx.doi.org/10.1023/A:1026670816596}{{\em Int. J. Theor. Phys.}
  {\bf 38} (1999)  1253--1271}, \href{http://arxiv.org/abs/gr-qc/9812034}{{\tt
  arXiv:gr-qc/9812034}}.

\bibitem{hu1999notes}
{B. L. Hu, A. Raval, and S. Sinha}, ``Notes on black hole fluctuations and
  backreaction,'' in {\em Black holes, gravitational radiation and the
  universe}, pp.~103--120.
\newblock Springer, 1999.

\bibitem{Hensh:2022pmh}
S.~Hensh, S.~Liberati, and V.~Vitagliano, ``{Tidal heating in a Riemann-Cartan
  spacetime},'' \href{http://arxiv.org/abs/2208.14262}{{\tt arXiv:2208.14262
  [gr-qc]}}.

\end{thebibliography}\endgroup

\end{document}